\documentclass[twocolumn,showpacs,preprintnumbers,amsmath,amssymb,superscriptaddress]{revtex4}

\usepackage{graphicx}
\usepackage{dcolumn}
\usepackage{bm}
\usepackage{tabularx}
\usepackage{ulem} 
\newcolumntype{M}{>{\centering\arraybackslash}m{1.85cm}}
\usepackage[export]{adjustbox}
\usepackage{float}
\usepackage{color}   
\usepackage{hyperref}
\hypersetup{
	colorlinks=true, 
	linktoc=all,     
	linkcolor=blue,  
}

\makeatletter
\newcommand{\colorcaption}[2][]{%
	\begingroup%
	\renewcommand{\@caption@fignum@sep}{ (Color online). }%
	\caption[#1]{#2}%
	\endgroup%
}
\makeatother
\begin{document}

\title{Systematic shell-model study of Rn isotopes with $A=$ 207 to 216 and isomeric states}

\author{Bharti Bhoy}
\email{bbhoy@ph.iitr.ac.in} 
\address{ Department of Physics, Indian Institute of Technology Roorkee, Roorkee
247 667, India}
\author{Praveen C. Srivastava}
\email{praveen.srivastava@ph.iitr.ac.in}
\address{ Department of Physics, Indian Institute of Technology Roorkee, Roorkee
247 667, India}

\date{\hfill \today}

\begin{abstract}

{\color{black} We present }systematic large-scale shell-model calculations for Rn isotopes with $A=$ 207 to 216. For the $^{207-212}$Rn isotopes, {\color{black} we perform } calculations with
KHH7B interaction, while for $^{213-216}$Rn isotopes with KHPE and KHH7B interactions. 
The calculated energies and electromagnetic properties {\color{black} are compared } with the available experimental data {\color{black} and predicted where experimental data are not available}. 
{\color{black} We also suggest } spins and parities of several unconfirmed states available from the recent experimental data. 
Comprehensive study of several isomeric states from the calculated shell-model configurations and half-lives is also reported. 

\end{abstract}

\maketitle

\section{INTRODUCTION}\label{1}

{\color{black}
 In the recent past, several experimental measurements have been done to study the structure and collectivity in the Pb region
 \cite{Brown2000, Butler, T.Otsuka, yosi, Tang,210At,210Ra,Ra196,210Po,208Fr,206Bi,Discovery1,Discovery2,prgati,berry, abinitio}.
 Apart from this, different class of correlations such as pairing, quadrupole and octupole ones are recently reported \cite{Butlerjpg}. 
   {\color{black} Nuclei in the vicinity of $N=126$ are very crucial to understand the
astrophysical $r$ process in producing nuclei heavier than A $\sim$ 190 \cite{Tang}.} A solvable model for octupole phonons for $^{208}$Pb is reported in Ref. \cite{isacker}.
   Also, several isomeric states are observed in this region \cite{jain,astier,203Tl,phil}.
   The level scheme of $^{212}$Rn 
 with up to spins of $\sim 39 \hbar$ and excitation energies of about 13 MeV has recently been reported using $^{204}$Hg($^{13}$C,5n)$^{212}$Rn reaction \cite{212Rn1}. In this experiment, two new isomers with $\tau =25(2)ns$ and $\tau = 12(2)ns$ were placed at 12.211 and 12.548 MeV, respectively. Theoretical results obtained by using semiempirical shell-model and  deformed independent particle model (DIPM) are also reported. In another recent experiment at Legnaro National Laboratory in Italy \cite{212Rn2}, a low-lying level scheme of  $^{212}$Rn has been populated. In this experiment, several non-yrast states based on $\pi h_{9/2}^4$ and $\pi h_{9/2}^3f_{7/2}$ configurations have been identified. Also, a $3^{(-)}$ collective state at 2.121 MeV is proposed. This state is believed to be arising from  mixing of the octupole vibration with a $3^-$ member of the $\pi h_{9/2}^3i_{13/2}$ multiplet.

 There have been several theoretical studies done in this mass region \cite{Mcgrory,Coraggio, Caurier, koji, Teruya, Yanase, Naidja, Wilson, Wahid, Anil}.
  McGrory and Kuo \cite{Mcgrory} have  reported the structure of the nuclei $^{204-206}$Pb, $^{210-212}$Pb, $^{210}$Po, $^{211}$At, and $^{212}$Rn with few valence nucleons away from the $^{204}$Pb core using conventional nuclear shell-model several decades before.
 However, with the increase in the computational facility, it becomes feasible to perform shell-model calculations for
nuclei having more valence nucleons. 
 Shell-model results using a realistic effective interaction, derived from the Bonn-A nucleon-nucleon potential by using a $G$-matrix folded-diagram approach for $N=126$ isotones are reported by Coraggio {\it et al.} in Ref. \cite{Coraggio}.  The Strasbourg group has reported large-scale shell model results for Po-Pu with $N=126$ using Kuo-Herling interaction in Ref. \cite{Caurier}.
 Yoshinaga group has reported shell model results for nuclei around mass 210 \cite{Teruya} and masses from 210 to 217 \cite{Yanase} using effective interaction with an extended pairing plus quadrupole-quadrupole interaction.

	
 
 The objective of the present study is to perform comprehensive shell-model calculations of $^{207-216}$Rn isotopes to cover nuclei below and above $N=126$ shell gap. 
 {\color{black} There is no systematic shell-model study available in the literature. From our shell-model study in two different 
model spaces, we have predicted the importance of different orbitals required to explain the high-spin states.
Since there are several new experimental data available for high spin states and corresponding isomers,
thus in our study, we have focused on high spin states.
Our results also confirm several unconfirmed states from the experiment. }
 The energy spectrum and electromagnetic properties are calculated and compared with the available experimental data. Isomeric states and respective half-lives in terms of the shell-model configurations and seniority are also presented. 
 It is important to note that  for the  $^{212}$Rn shell-model results are available with the KHPE interaction \cite{212Rn2}. However, we have done calculations with the KHH7B interaction to see the importance of including lower orbitals in the model space. 
 
 The outline of the paper is as follows. In Sec. \ref{2}, the theoretical formalism of the present shell-model study is given. In Sec. \ref{3}, we present the results obtained for the energy spectrum, electromagnetic properties, and half-lives for isomers and compare them with the available experimental data. Sec. \ref{4} contains a summary and conclusions of the present work.

 \section{Formalism : SHELL-MODEL SPACE AND INTERACTIONS}\label{2}

Systematic studies have been carried out to understand the structure of Rn isotopes with $A$ = 207-216 considering two different sets of interactions and valence spaces. To diagonalize the matrices, the NUSHELLX \cite{Nushellx1, Nushellx2} and KSHELL \cite{Kshell}  codes have been employed for the shell-model calculations. Here we have taken two interactions for two different sets of isotopes, KHPE \cite{Warburton1}, and KHH7B \cite{pbpop}. Our focus is mainly on the application of KHH7B on the whole Rn chain considered.  To handle large dimensions, the KSHELL shell-model code is used. 
The highest dimension is 1.3 x 10$^9$  corresponding to ground state for $^{216}$Rn with the KHPE interaction.
For $A=213-216$, we have used KHPE interaction, and calculations using KHH7B interaction have also been done using NUSHELLX code with truncation in the model space.
Computationally it is  challenging to perform shell model calculations without truncation in the Pb region.

One of the interactions we are using in our calculation is KHPE. The model space here consists of $1h_{9/2}, 2f_{7/2}, 2f_{5/2}, 3p_{3/2}, 3p_{1/2}, 1i_{13/2}$ proton orbitals and  $1i_{11/2}, 2g_{9/2}, 2g_{7/2}, 3d_{5/2}, 3d_{3/2}, 4s_{1/2}, 1j_{15/2}$ neutron orbitals. 
The KHH7B residual interaction used by Poppelier and Glaudemans \cite{pbpop} is the Surface Delta Interaction (SDI), which is the schematic interaction but gives the same results as the Kuo-Herling matrix elements \cite{Kuo1, Kuo2}.
The effective realistic residual interaction of Kuo and Herling \cite{Kuo1, Kuo2} was derived from a free nucleon-nucleon potential of Hamada and Johnston \cite{Hamada} with renormalization due to the finite extension of model space by the reaction matrix techniques developed by Kuo and Brown \cite{Kuo3}. 
{\color{black}In the present work, we have performed shell model calculations with KHPE interaction without any truncations. The full-fledged calculation with KHPE interaction is sufficient to explain low-lying states,  but we need core-excitation for the explanation of high-lying states.}

The KHH7B interaction consists of the four proton orbitals $2d_{5/2}, 2d_{3/2}, 3s_{1/2}, 1h_{11/2}$ below and three orbitals $1h_{9/2}, 2f_{7/2}, 1i_{13/2}$ above the $Z$ = 82, and four neutron orbitals  $2f_{5/2}, 3p_{3/2},  3p_{1/2}, 1i_{13/2}$ below and three orbitals $2g_{9/2}, 1i_{11/2}, 1j_{15/2}$ above $N$ = 126 energy gap.
In KHH7B interaction, the cross shell two-body matrix elements (TBMEs) were
generated by the G-matrix potential (H7B) \cite{Hosaka}, while the proton-neutron, hole-hole, and particle-particle TBMEs are taken from Kuo-Herling interaction \cite{Kuo1} as modified in the Ref. \cite{Warburton1}.
Previously, shell model results using KHH7B interaction are reported in \cite{Wilson,berry, Wahid,Anil}.
{\color{black}For KHH7B, we have completely filled proton orbitals below $Z = 82$, while neutrons are only allowed to occupy the orbitals below $N=126$  for $A = 207-212$ and  above $N=126$ for $A = 213-216$. 
In the Pb region, shell-model calculation taking into account core-excitation is very important. However, due to huge-dimension, we are unable to perform an appropriate calculation using core-excitation.  It is essential to perform the shell-model calculation using KHH7B interaction without any truncation to see the role of lower orbitals.
Further, we need to develop a new interaction, because existing interactions are very old. Although, we have limited experimental data to tune the effective interaction. In the Pb region, new data for  both energy and electromagnetic properties are coming. In the future, it is possible to construct a new interaction with these data.}

\begin{figure}
	 \includegraphics[width=85mm,height=105mm]{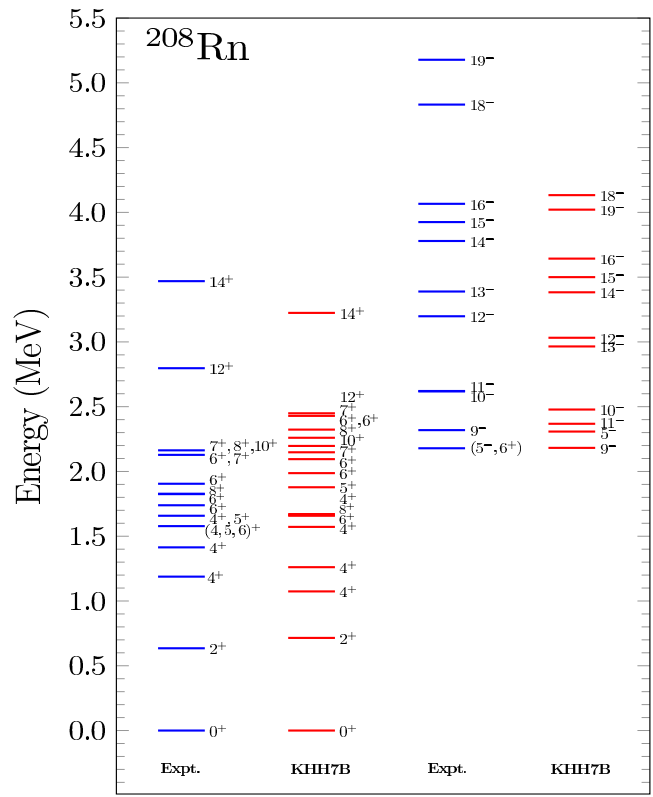}
	
	\caption{\label{fig1} Comparison between calculated and experimental \cite{NNDC} energy levels for $^{208}$Rn.}
\end{figure}

\begin{figure}
	\includegraphics[width=85mm,height=105mm]{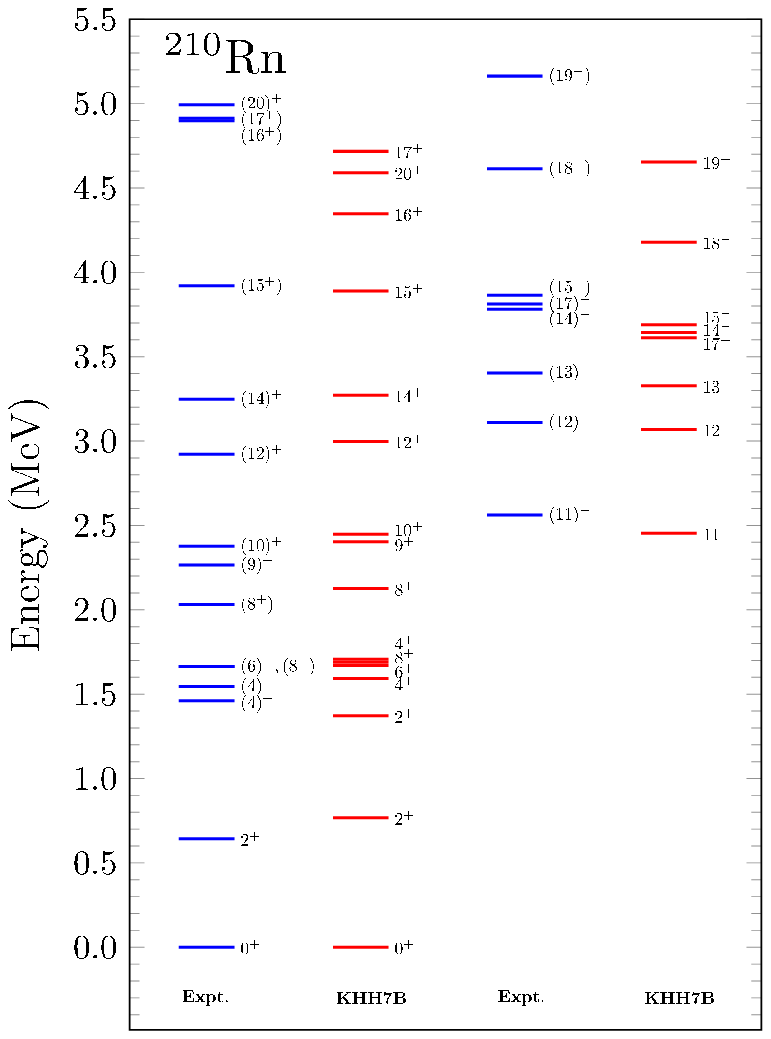}

	\caption{\label{fig2} Comparison between calculated and experimental \cite{NNDC} energy levels for $^{210}$Rn.}
\end{figure}

\begin{figure}
	\includegraphics[width=85mm,height=105mm]{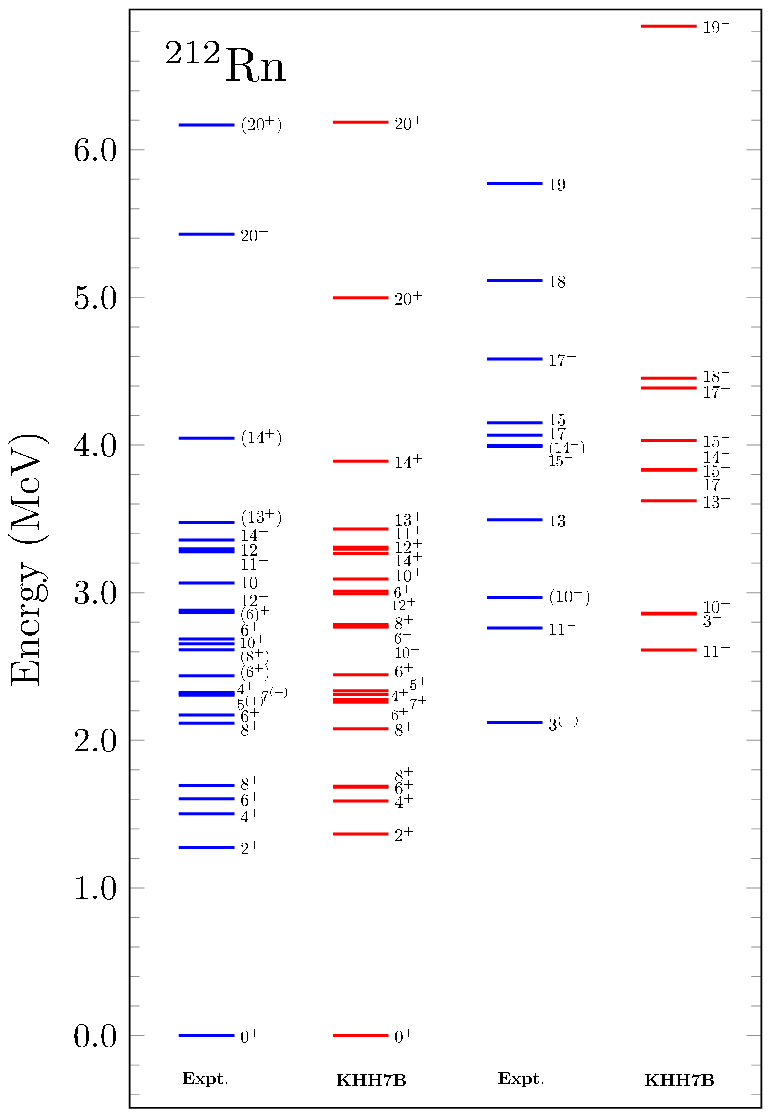}
		
	\caption{\label{fig3} Comparison between calculated and experimental \cite{NNDC}, \cite{212Rn2} energy levels for $^{212}$Rn.}
\end{figure}

\begin{figure}
	\includegraphics[width=85mm,height=105mm]{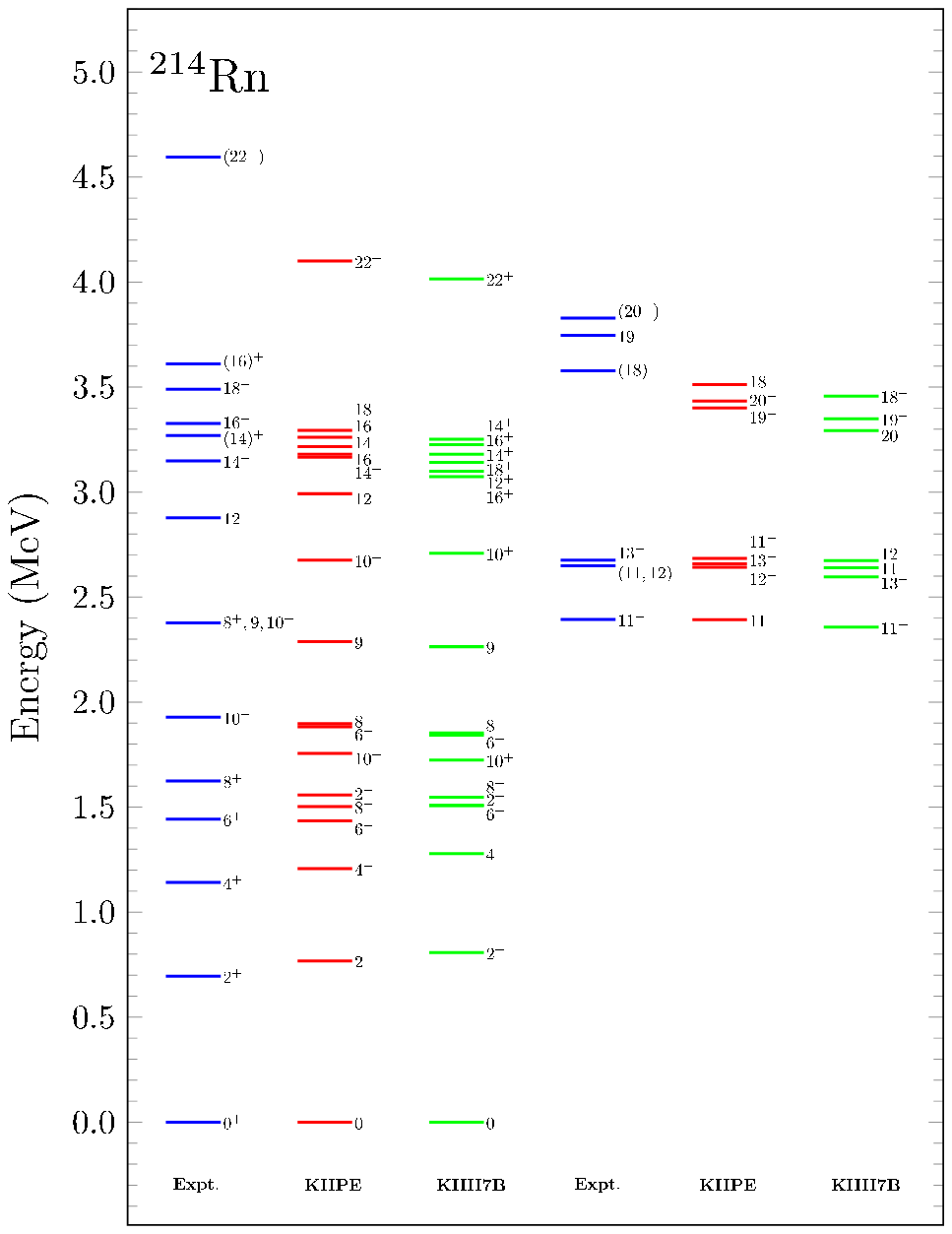}
	\caption{\label{fig4} Comparison between calculated and experimental \cite{NNDC} energy levels for $^{214}$Rn.}
\end{figure}

\begin{figure}
	\includegraphics[width=85mm,height=105mm]{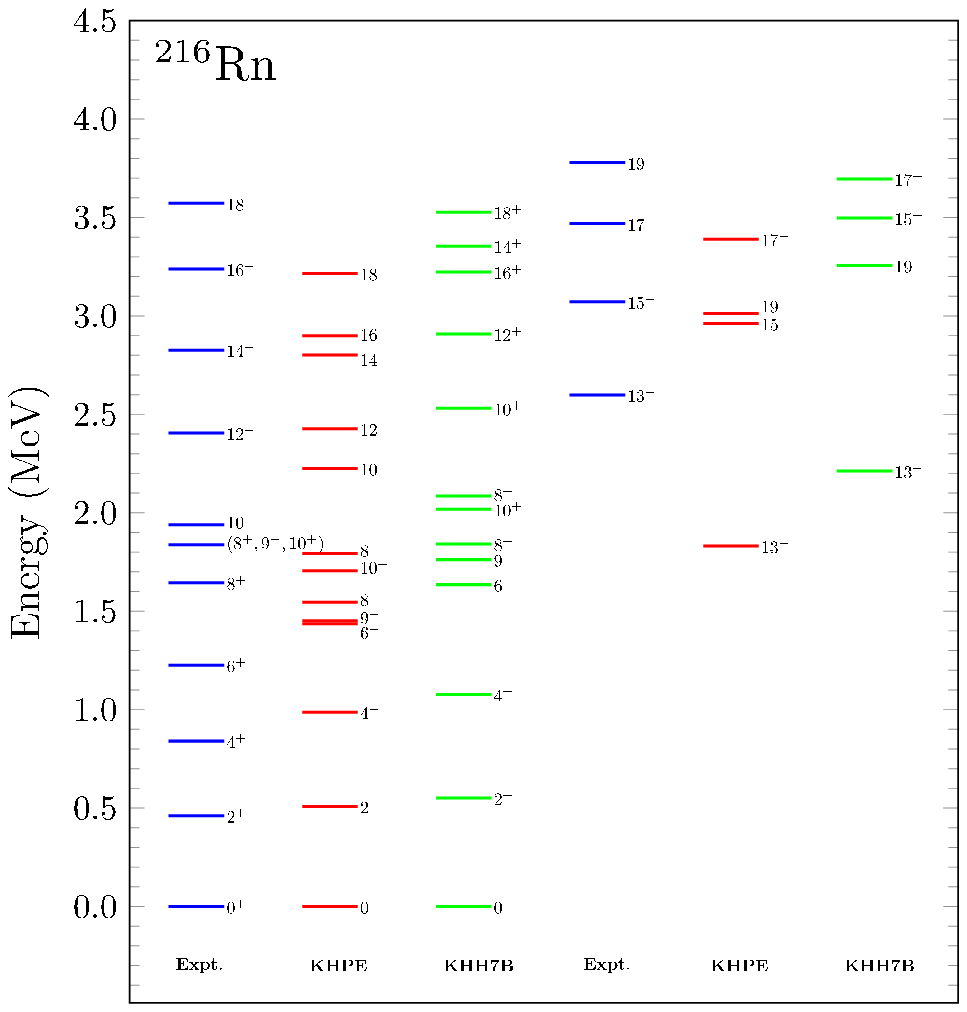}

	\caption{\label{fig5} Comparison between calculated and experimental \cite{NNDC} energy levels for $^{216}$Rn.}
\end{figure}

\section{RESULTS AND DISCUSSION}\label{3}

In this section, the results of our calculations for the Rn isotopes are presented and compared with experimental data. 
In Figs. \ref{fig1} to \ref{fig5} results obtained for the 
$^{208,210,212,214,216}$Rn isotopes are shown in comparison with experimental data. Results for the  $^{207,209,211,213,215}$Rn isotopes are given in Figs. \ref{fig6} to \ref{fig10}.
The configuration of the probability for dominant wave-function and corresponding seniority of isomeric states are reported in Table \ref{t_Con}. The electromagnetic properties are reported in
Tables \ref{t_be2} and \ref{t_Q}. 
 The calculated half-lives corresponding to the isomeric states are reported in comparison with the experimental data in Table \ref{t_hl}.

\subsection{Even Rn isotopes}

 Fig. \ref{fig1} shows the shell-model energy spectrum of $^{208}$Rn in comparison with the experimental data, where all the observed levels up to 5.2 MeV excitation energy are reported. In the calculated low-energy spectrum, we get a much higher level density compared to experimental data. We report only those yrast and non-yrast shell-model states corresponding to the experimental levels. 
The shell-model is giving close energy values for energy levels up to 2.5 MeV for all the yrast states.
In our calculation, the experimental ${(4,5,6)^+}$ level at 1.580 MeV can be associated with either the $4_3^+$ or $6_1^+$ state. The calculated $5_1^+$ state is produced 300 keV higher than  the experimental state at 1.580 MeV, and calculated $5_2^+$ is produced 500 keV higher than the experimental state at 1.658 MeV ( ${4}^+$, ${5}^+$). Therefore, we propose the assignment of calculated $5_1^+$ state to the closest experimental level at 1.658 MeV, excluding ${5}^+$ from the 1.580 MeV experimental state.
For the negative parity state,
the experimental ${(5^-,6^+)}$ level at 2.18 MeV can be associated with the calculated $5_1^-$ at 2.308 MeV.

The shell-model energy spectrum of $^{210}$Rn in comparison with the experimental data is shown in Fig. \ref{fig2}, where all the observed levels up to 5.2 MeV excitation energy are reported. Experimentally, the relative energies between the ground state and many excited states are unknown in $^{210}$Rn. The $(8_1^+)$ state of these excited states is the lowest and observed at 1.665+x MeV. In this figure, these states are shown without any assumption of x ( i.e., x=0). The shell-model reproduces the energy levels well up to 4 MeV for both positive and negative parity states. Most of the levels in the case of $^{210}$Rn are tentative. 
Therefore, a one-to-one correspondence has been estimated with the experimental data. 
The second excited state in the calculation is produced as ${2}^+$, 90 keV lower than the experimental energy value, which is an experimentally tentative $({4})^+$ state. 
The calculated $4_1^+$ state is produced at 1.593 MeV close to the  tentative experimental   level $({4})^+$ at 1.545 MeV.
The calculated $6_1^+$, $8_1^+$ and $8_2^+$ are confirming the  tentative experimental  states with a difference of a few keV.
{\color{black} The last two calculated negative parity states, $18_1^-$  and $19_1^-$ are lower in energies with respect to the experimental data by 436 keV and 572 keV, respectively. }
These levels might arise from  core-excitations, and in some cases, significant admixtures of configurations without and with  core-excitation beyond $Z=82$ and $N=126$ shell closure.

The shell-model energy spectrum of $^{212}$Rn in comparison with the experimental data is shown in Fig. \ref{fig3},  where all the observed levels up to 6.2 MeV excitation energy are reported. 
In this work, we have reproduced all the new states identified in \cite{212Rn2}, with a satisfactory quantitative agreement between our results and experimental data. 
A  direct one-to-one correspondence  between states can be established up to 2.300 MeV in positive parity states.
Here we observe similar trend in $^{208}$Rn, $^{210}$Rn and $^{212}$Rn isotopes: a small energy gap  between the ${6_1}^+$ and ${8_1}^+$ states, and a large gap between $8_1^+$ and $8_2^+$ states  are featured.
Our calculation confirms the  tentative experimental  $6_3^+$ and $13_1^+$ states with a difference of only a few keV. 
The calculated $6_5^+$ and $12_1^+$ states are 111 keV and 144 keV higher than the experimental data, respectively.  This state could be ${6_5}^+$ state.  
For high-lying states, the compression becomes notable, except for the calculated $20_2^+$ state and confirms this tentative state. 
The $18_1^-$ state is 662 keV lower, and $19_1^-$ state is 1.065 MeV higher than the experimental data.   The description of these high-lying states requires the core-excitation  and maybe some mixing between the single-particle states and  core-excitation above $Z=82$. 
and $N=126$ shell closure. 
In our calculation, the collective $3_1^-$ state at 2.856 MeV is 735 keV higher than the proposed experimental value, and arising from the configuration $\pi(h_{9/2})^3(i_{13/2})$.
The single-particle orbitals with $\Delta l$ = 3, $\pi{f_{7/2}}-\pi{i_{13/2}}$ and $\nu{g_{9/2}}-\nu{j_{15/2}}$ in our model space and core-excitations across the two shell gaps at $Z=82$ and $N=126$ are responsible for the structure of this octupole vibration.
The KHPE (results as reported in Ref. \cite{212Rn2}) and KHH7B interactions are giving almost similar results because we have only four valence protons beyond Z=82 and N=126. This is reflected from the  similar wave functions we have obtained from these two interactions.

The shell-model energy spectrum of $^{214}$Rn in comparison with the experimental data is shown in Fig. \ref{fig4} with the two interactions KHPE and KHH7B,  where all the observed levels up to 4.6 MeV excitation energy are reported. 
In $^{214}$Rn, many of the energy levels are unidentified in terms of the spin-parity. These levels are not included in our figure. However, we have tried to interpret the first two unidentified levels. The shell-model results from the KHPE interaction are slightly better than the KHH7B interaction, as shown in Fig. \ref{fig4}.
Similar to other even Rn isotopes, a small energy gap is observed between the $6_1^+$ and $8_1^+$ states, and a large gap between $8_1^+$ and $8_2^+$ states.
 The spin-parities of the experimental states at 1.332 MeV and 1.800 MeV are not assigned. In the experimental data, the first state decays to the $2_1^+$ state at 0.695 MeV, and the second state decays to the first unidentified state at 1.332 MeV.
The calculated $2_2^+$ state is at 1.558 MeV and 1.509 MeV from the KHPE and KHH7B interaction, respectively.
The calculated $6_2^+$  state is at 1.883 MeV and 1.842 MeV from the KHPE and KHH7B interaction, respectively. 
From the comparison with neighboring nuclei $^{212}$Rn, $^{216}$Rn from both our calculation and experimental data, and considering our results for the above two mentioned states, we suggest the spin-parities of the experimental state at 1.332 MeV to be $2_2^+$ and state at 1.800 MeV to be $6_2^+$. 
With both interactions the calculated $12_1^-$ state is very close to the tentative experimental state with only few keV differences.
The calculated negative parity states above 3.0 MeV are highly compressed.
The $^{214}$Rn isotope is above the $Z=82$ and $N=126$ shell-closure with 4 valence protons and 2 valence neutrons. Therefore, it is important to consider sufficient orbitals around the shell-closure for  core-excitation and required configuration mixing.
Due to limited computational facilities, we are not able to include all orbitals without truncation.

The shell-model energy spectrum of $^{216}$Rn in comparison with the experimental data is shown in Fig. \ref{fig5} with two interactions KHPE and KHH7B up to 3.8 MeV excitation energy. 
{\color{black} Previously observed feature for a small energy gap between the $6_1^+$ and $8_1^+$ states, and a large gap between $8_1^+$ and $8_2^+$ states, are not observed here as prominently as in the other even Rn isotopes. 
However, } as observed in the experiment, this pattern has vanished in $^{216}$Rn due to the enhancement of quadrupole collectivity.  
As we move from $^{214}$Rn to $^{216}$Rn, the number of valence particles increases beyond $Z=82$ and $N=126$ shell closure, thus we need to include more orbitals in the model space apart from the  core-excitation.
Therefore, both the interactions are not being able to reproduce the levels in the whole energy range.
Due to the model space requirement for any specific state, some levels are in good agreement with experimental data, while others could not be reproduced well. 
 The spectrum of $^{216}$Rn looks like vibrational one because low-lying states from $0^+$ - $8^+$ are equally spaced
 . In the case of vibrational spectra, the quadrupole moment
should be smaller. In the simplest version of the vibrational model, the quadrupole moment of the $2^+$ state is predicted to be zero.
Our shell model results give an equal spacing of low-lying states,  although  results for quadrupole-moments are large.
{\color{black} However,} there is no experimental data available for the quadrupole moment.

\begin{figure}
	\includegraphics[width=85mm,height=105mm]{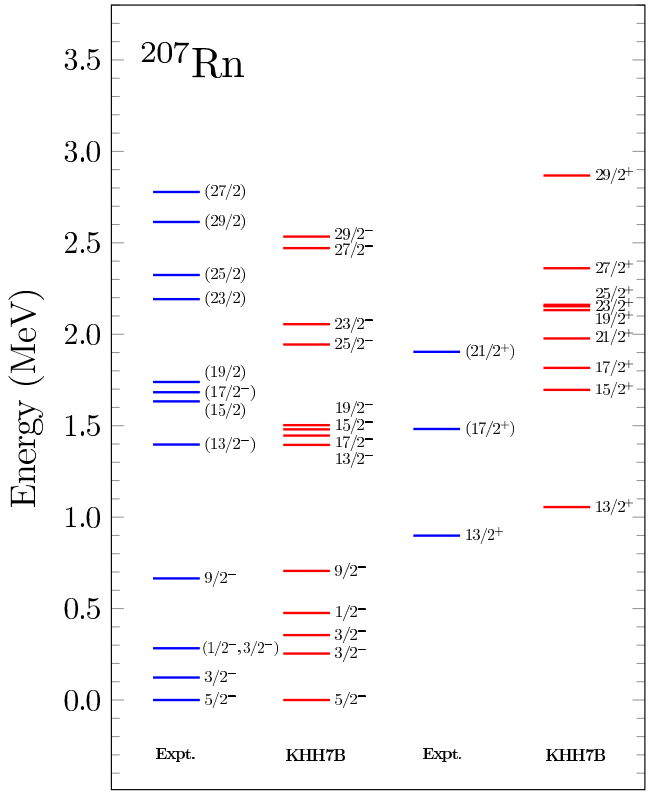}
	
	\caption{\label{fig6} Comparison between calculated and experimental \cite{NNDC} energy levels for $^{207}$Rn.}
\end{figure}

\begin{figure}
	\includegraphics[width=85mm,height=105mm]{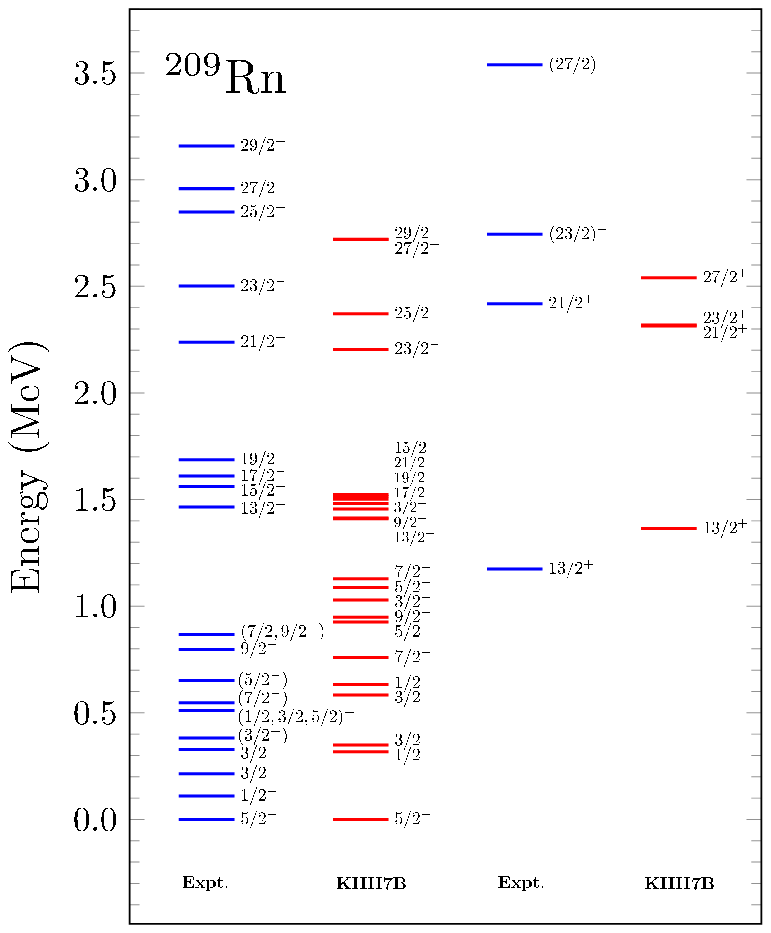}

	\caption{\label{fig7} Comparison between calculated and experimental \cite{NNDC} energy levels for $^{209}$Rn.}
\end{figure}

\begin{figure}
	\includegraphics[width=85mm,height=105mm]{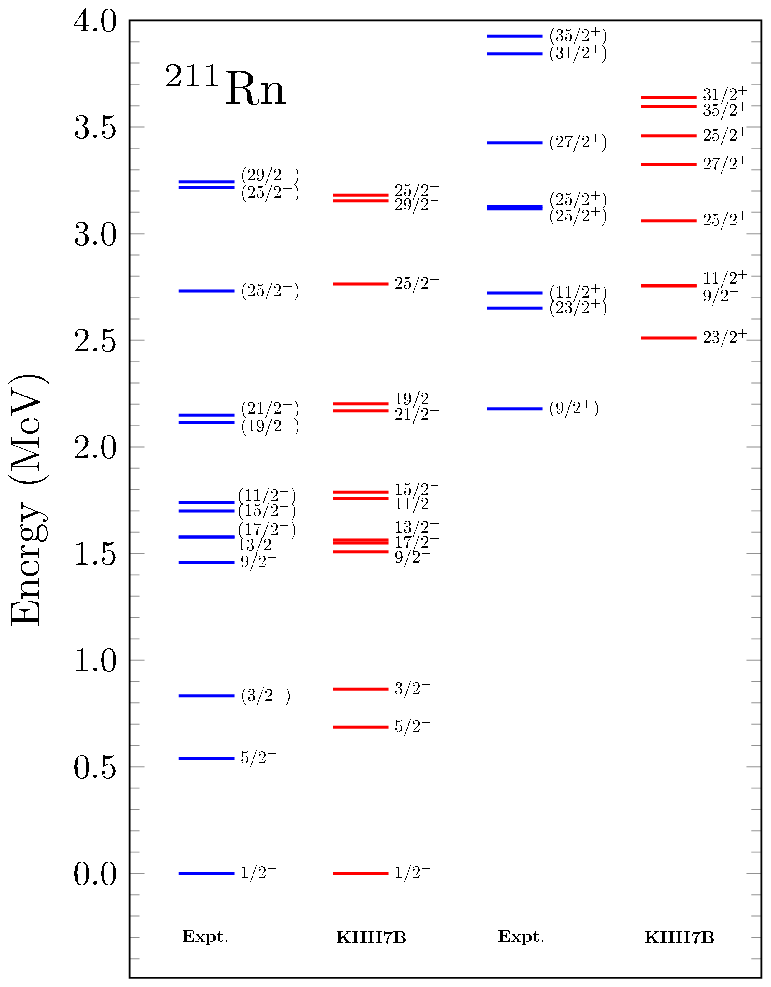}

	\caption{\label{fig8} Comparison between calculated and experimental \cite{NNDC} energy levels for $^{211}$Rn.}
\end{figure}

\begin{figure}
	\includegraphics[width=85mm,height=105mm]{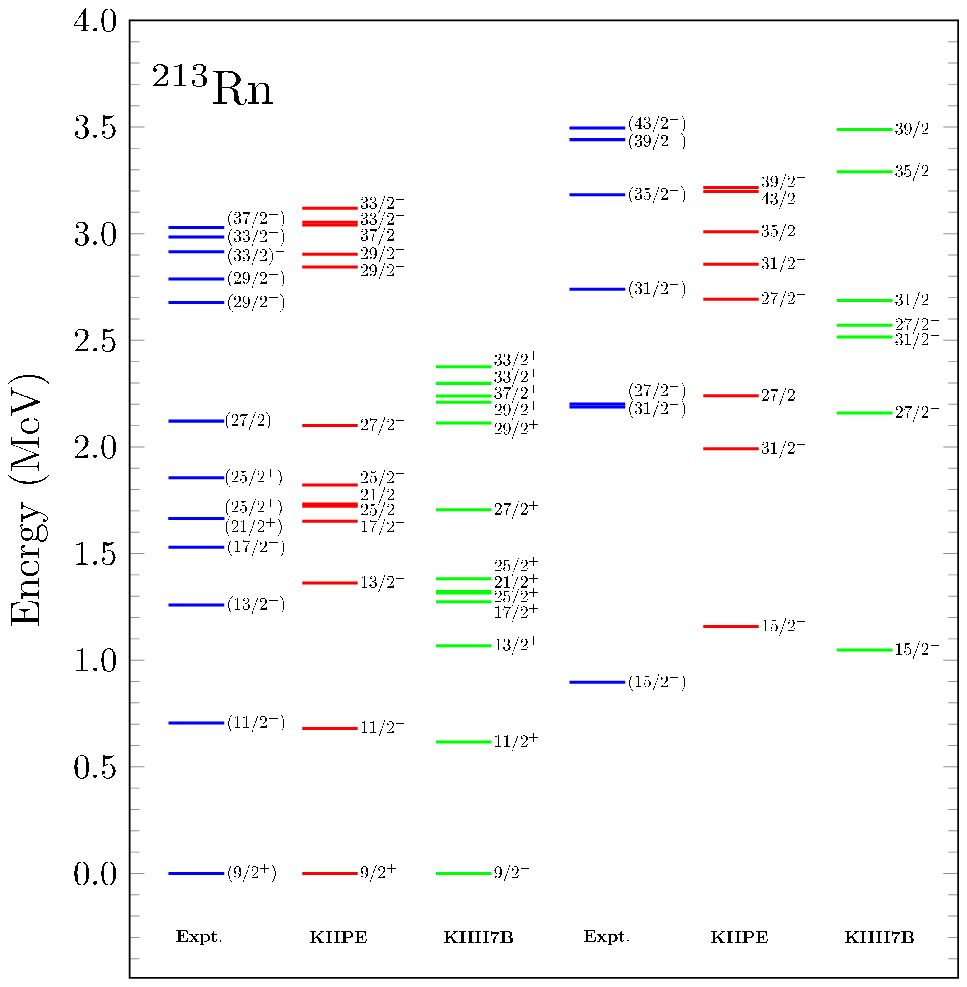}

	\caption{\label{fig9} Comparison between calculated and experimental \cite{NNDC} energy levels for $^{213}$Rn.}
\end{figure}

\begin{figure}
	\includegraphics[width=85mm,height=105mm]{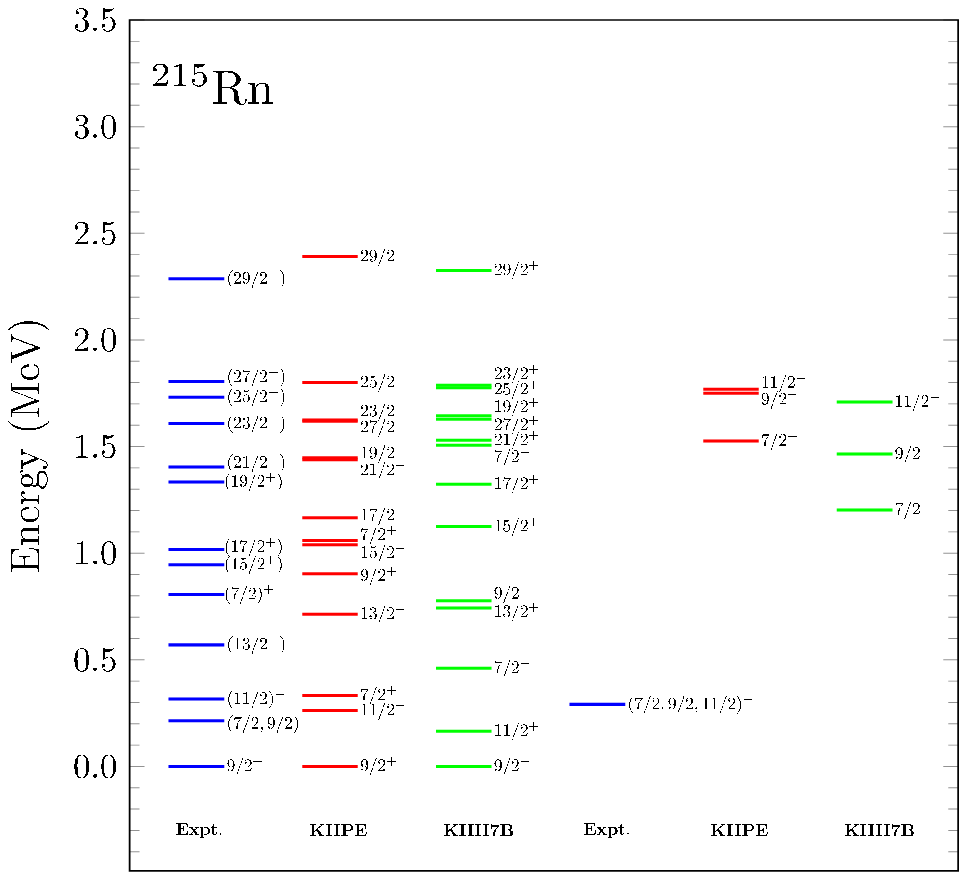}

	\caption{\label{fig10} Comparison between calculated and experimental \cite{NNDC} energy levels for $^{215}$Rn.}
\end{figure}

\begin{table}
	\begin{center}
		\caption{Configurations of isomeric states in Rn isotopes with the probability of the dominant component of the configuration.}
		\label{t_Con}
		
		\begin{tabular}{r|r|c|c|cc}
			\hline 
			\hline	
			&  &  &   & &   \\
			Nucleus & $J^{\pi}$ & Seniority  & Wave-function & Probability  \\
			
			&  &  &   & &   \\
			\hline
			
			&  &  &   & &   \\
			$^{207}$Rn  & $13/2_1^+$ & $v=1$ & $\nu(i_{13/2})^{-1}$ & 25.28$\%$    \\\hline

			&  &  &   & &   \\
			$^{209}$Rn  & $13/2_1^+$ & $v=1$ & $\nu(i_{13/2})^{-1}$ & 26.33$\%$    \\
			& $29/2_1^-$ & $v=3$ & $\pi(h_{9/2})^4\otimes\nu(f_{5/2})^{-1}$ & 73.43$\%$    \\
			&$35/2_1^+$ & $v=5$ & $\pi(h_{9/2})^3(i_{13/2})\otimes\nu(f_{5/2})^{-1}$ & 58.53$\%$   \\
			& & & $\pi(h_{9/2})^3(i_{13/2})\otimes\nu(p_{1/2})^{-1}$ & 11.45$\%$  \\
			& $41/2_1^-$ & $v=5$ & $\pi(h_{9/2})^2(i_{13/2})^2\otimes\nu(f_{5/2})^{-1}$ & 49.48$\%$    \\
			&  & & $\pi(h_{9/2})^2(i_{13/2})^2\otimes\nu(p_{1/2})^{-1}$ & 17.27$\%$ \\
			\hline

			&  &  &   & &   \\
			$^{211}$Rn  & $17/2_1^-$ & $v=3$  & $\pi(h_{9/2})^4\otimes\nu(p_{1/2})^{-1}$ & 69.04$\%$    \\
			& $35/2_1^+$ & $v=5$ & $\pi(h_{9/2})^3(i_{13/2})\otimes\nu(p_{1/2})^{-1}$ & 85.45$\%$    \\
			& $43/2_1^-$ & $v=5$ & $\pi(h_{9/2})^2(i_{13/2})^2\otimes\nu(f_{5/2})^{-1}$ & 91.30$\%$   \\
			& $49/2_1^+$ & $v=5$ & $\pi(h_{9/2})^2(i_{13/2})^2\otimes\nu(g_{9/2})$ & 61.78$\%$    \\
			& $63/2_1^-$ &  $v=7$ & $\pi(h_{9/2})^2(i_{13/2})^2\otimes$ & 88.04$\%$ \\
			&  &   & $\nu(g_{9/2})(i_{11/2})(f_{5/2})^{-1}$ &  \\
			\hline 
			
			&  &  &   & &   \\
			$^{213}$Rn  & $15/2_1^-$ & $v=1$ & $\nu(j_{15/2})$ & 41.94$\%$    \\
			& $21/2_1^+$ & $v=3$ & $\pi(h_{9/2})^4\otimes\nu(g_{9/2})$ & 51.79$\%$    \\
			& $25/2_1^+$ & $v=3$ & $\pi(h_{9/2})^4\otimes\nu(g_{9/2})$ & 48.37$\%$   \\
			& $31/2_1^-$ & $v=3$ &$\pi(h_{9/2})^3(i_{13/2})\otimes\nu(g_{9/2})$ & 63.64$\%$  \\
			& $37/2_1^+$ & $v=5$ & $\pi(h_{9/2})^3(f_{7/2})\otimes\nu(g_{9/2})$ &95.82$\%$    \\
			& $43/2_1^-$ & $v=5$ & $\pi(h_{9/2})^3(i_{13/2})\otimes\nu(g_{9/2})$ & 92.94$\%$ \\
			& $49/2_1^+$ & $v=5$ & $\pi(h_{9/2})^2(i_{13/2})^2\otimes\nu(g_{9/2})$  & 91.62$\%$    \\
			& $55/2_1^+$ & $v=5$ & $\pi(h_{9/2})(i_{13/2})^3\otimes\nu(j_{15/2})$ & 99.95$\%$    \\
			\hline

			&  &  &   & &   \\
			$^{208}$Rn  & $8_1^+$ & $v=2$ & $\pi(h_{9/2})^4$ & 38.74$\%$    \\
			& $10_1^-$ &  $v=2$ & $\pi(h_{9/2})^3(i_{13/2})$ & 32.88$\%$    \\
			& $16_1^-$ & $v=2$ & $\pi(h_{9/2})^3(i_{13/2})$ & 43.65$\%$   \\
			\hline

			&  &  &   & &   \\
			$^{210}$Rn  & $8_1^+$ & $v=2$ & $\pi(h_{9/2})^4$ & 46.63$\%$    \\
			& $11_1^-$ &$v=2$   & $\pi(h_{9/2})^3(i_{13/2})$ & 42.72$\%$    \\
			& $14_1^+$ & $v=4$ & $\pi(h_{9/2})^3(f_{7/2})$ & 49.23$\%$   \\
			& $17_1^-$ &  $v=4$ & $\pi(h_{9/2})^3(i_{13/2})$ & 50.80$\%$    \\
			& $20_1^+$ &$v=4$  & $\pi(h_{9/2})^2(i_{13/2})^2$ & 42.73$\%$   \\
			\hline

			&  &  &   & &   \\
			$^{212}$Rn  & $6_1^+$ & $v=2$ & $\pi(h_{9/2})^4$ & 67.34$\%$    \\
			& $8_1^+$ & $v=2$ & $\pi(h_{9/2})^4$ & 68.42$\%$    \\
			&  $14_1^+$  & $v=4$ & $\pi(h_{9/2})^3(f_{7/2})$ & 98$\%$ \\
			& $17_1^-$ & $v=4$  & $\pi(h_{9/2})^3(i_{13/2})$ & 99.59$\%$    \\
			& $22_1^+$ &$v=6$ & $\pi(h_{9/2})^3(i_{13/2})\otimes$ & 80.48$\%$   \\
			&  & & $\nu(p_{1/2})^{-1}(g_{9/2})$ &    \\
			& $25_1^-$ & $v=6$ & $\pi(h_{9/2})^2(i_{13/2})^2\otimes$ & 73.45$\%$    \\
			& & & $\nu(p_{1/2})^{-1}(g_{9/2})$ &    \\
			\hline

			&  &  &   & &   \\
			$^{214}$Rn   & $18_1^+$ & $v=4$ & $\pi(h_{9/2})^3(i_{13/2})\otimes\nu(g_{9/2})^2$ & 60$\%$   \\
			& $22_1^+$ & $v=4$& $\pi(h_{9/2})^3(i_{13/2})\otimes$ & 83$\%$   \\
			&  & & $\nu(g_{9/2})(j_{15/2})$ &    \\
			\hline
			
		\end{tabular}
	\end{center}
\end{table}

The seniority, configurations and half-life corresponding to different isomeric states for Rn isotopes as reported in  Table \ref{t_Con} and
Table \ref{t_hl} are discussed below. 
{\color{black} In spherical nuclei close to the magic number, the breaking of high-$j$ nucleon pairs produces isomeric states. The Rn isotopes considered in our calculations are spherical, thus it is possible to explain the isomers in terms of seniority quantum number.
  Seniority $\nu$ is the number of particles not in pairs coupled to angular momentum $J$ = 0.
  With shell-model, it is possible to extract information 
 about seniority from the configurations. 
 For the seniority isomer, the decay is hindered because of the same seniority of the initial and final states.}
 Seniority isomers arise because $E2$ decay is hindered between the same seniority of the initial and final states.
  In  the case of $^{208,210,212}$Rn isotopes, $8^+$ state is the seniority isomer with a very small $B(E2$; $8^+$ $\rightarrow$ $6^+$) value. This is because
of the same seniority $v=2$ for $8^+$ and $6^+$, which is coming from $\pi(h_{9/2})^4$ configuration.

In $^{208}$Rn, $^{210}$Rn and $^{212}$Rn the $8_1^+$ isomeric state is coming from $\pi(h_{9/2})^4$ configuration, in increasing probability with mass number. The $8_1^+$ isomeric state is formed by purely $h_{9/2}$ orbital with one pair breaking, thus $\nu$ = 2. The half-life of this isomeric state is also reproduced very close to the experimental data with $B(E2)$ transition for all three isotopes. The $10_1^-$ and $16_1^-$ states are coming from the same configuration $\pi(h_{9/2})^3(i_{13/2})$, and formed by one proton in $h_{9/2}$, $i_{13/2}$  each coupled with one pair in $h_{9/2}$ orbital, hence the seniority $\nu$ = 2. These two states are generated by $B(M1)$ transition and half-lives for these states are reproduced well with the experimental data. In $^{210}$Rn, $11_1^-$[$\pi(h_{9/2})^3(i_{13/2})$] isomeric state is formed by one proton in $h_{9/2}$ and $i_{13/2}$ orbitals each, thus $\nu$ = 2. 
The ${14_1}^+$[$\pi(h_{9/2})^3(f_{7/2})$] isomeric state is formed by one pair breaking in $\pi(h_{9/2})$ orbital, and one proton in $h_{9/2}$ and $f_{7/2}$ orbitals each, thus $\nu$ = 4. The half-life for this state is reproduced in the same order as the experimental data 76(7) ns,  with a higher magnitude, as the $B(E2)$ transition value in our calculation is smaller than the experimental data. The ${17_1}^-$[$\pi(h_{9/2})^3(i_{13/2})$] isomeric state is formed by  one pair breaking in $\pi(h_{9/2})$ orbital, and one proton in $h_{9/2}$ and $i_{13/2}$ orbitals each, thus $\nu$ = 4. The $20_1^+$[$\pi(h_{9/2})^2(i_{13/2})^2$] isomeric state is formed by one pair breaking in $\pi(h_{9/2})$ and $\pi(i_{13/2})$ both the orbitals, thus $\nu$ = 4. 
In $^{212}$Rn, the $6_1^+$ isomeric state is formed similar to the $8_1^+$ states in all even isotopes with the same seniority. The half-life of this state is almost the same as the experimental data, as our $B(E2)$ transition value and energy difference are reproduced very well.
The $14_1^+$ and $17_1^-$ isomeric states are similar as in the $^{210}$Rn, with almost double probability (98$\%$ and 99.59$\%$, respectively) as these states are pure states. The half-lives for $14_1^+$ and $17_1^-$ isomeric states are coming in the same order with higher magnitude because the calculated $B(E2)$ transition value is significantly less than the experimental data for the $14_1^+$ state, and our energy difference is very small for $17_1^-$. The $22_1^+$[$\pi(h_{9/2})^3(i_{13/2})\otimes\nu(p_{1/2})^{-1}(g_{9/2})$] isomeric state is formed by one pair breaking in $\pi(h_{9/2})$ orbital, and one proton in $h_{9/2}$ and $i_{13/2}$ orbitals each, and one neutron in $p_{1/2}$ and $g_{9/2}$ orbitals each, thus $\nu$ = 6. The $25_1^-$[$\pi(h_{9/2})^2(i_{13/2})^2\otimes\nu(p_{1/2})^{-1}(g_{9/2})$] isomeric state is formed by one pair breaking in $\pi(h_{9/2})$ and $\pi(i_{13/2})$ both the orbitals and one neutron in $p_{1/2}$ and $g_{9/2}$ orbitals each, thus $\nu$ = 6. 
In $^{214}$Rn, the ${18_1}^+$[$\pi(h_{9/2})^3(i_{13/2})\otimes\nu(g_{9/2})^2$] isomeric state is formed by one pair breaking in $\pi(h_{9/2})$ or $\nu(g_{9/2})$ orbital, and one proton in  $h_{9/2}$ and $i_{13/2}$ orbitals each, thus $\nu$ = 4.
The $22_1^+$ isomeric state is arising from the configuration $\pi(h_{9/2})^3(i_{13/2})\otimes\nu(g_{9/2})(j_{15/2})$, and formed by one proton in $h_{9/2}$ and $i_{13/2}$ orbitals each, and one neutron in $g_{9/2}$ and $j_{15/2}$ orbitals each, thus $\nu$ = 4. For this isomeric state, seniority  reduces as compared to the $22_1^+$ state in $^{212}$Rn because of the role of $j_{15/2}$ orbital.  It is important to mention here that to calculate $49/2_1^+$ and  $63/2_1^-$ for $^{211}$Rn and $22_1^+$ and $25_1^-$ for $^{212}$Rn, we have opened $2g_{9/2}, 1i_{11/2}$, and $1j_{15/2}$ neutron orbitals, allowing two-particle excitation in each orbital (with restriction on other orbitals also). Corresponding wavefunctions are shown in Table \ref{t_Con}, and 
the calculations are performed using the NUSHELLX  code.

\subsection{Odd Rn isotopes}

Fig. \ref{fig6} shows the shell-model energy spectrum of $^{207}$Rn in comparison with the experimental data, where all the observed levels up to 2.8 MeV excitation energy are reported. 
 In $^{207}$Rn, most of the experimental levels above 1.5 MeV are not assigned with confirmed  spin-parities. Therefore, we have calculated the same levels for both parities to see a resemblance between the experimental data and shell-model. For the known experimental states with both spin-parity, the shell-model reproduces the energy spectrum with good agreement. The calculated $3/2_1^-$ state is twice the energy of the experimental data. {\color{black} On the other hand}, following the trend of our calculated spectrum, 100-200 keV range discrepancy in energy is negligible, and the experimental $3/2_1^-$ state is at a very small energy value of 0.123 MeV. The experimental state at 0.283 MeV can be associated with both $3/2_2^-$ and $1/2_1^-$ calculated states. 
Above 1.4 MeV, all the calculated states are compressed. 

We can see from  Fig. \ref{fig6}, our calculation can not predict precisely the unconfirmed experimental states, as they are equally likely  be assigned in both the parities.

The shell-model energy spectrum of $^{209}$Rn in comparison with the experimental data is shown in Fig. \ref{fig7},  where all the observed levels up to 3.6 MeV excitation energy are reported. In $^{209}$Rn, the energy spectrum is dense for low-lying states, and the calculated spectrum has reproduced the experimental data with a quantitative agreement. We have calculated all of the low-lying experimental levels, but we have excluded few non-yrast calculated levels from Fig. \ref{fig7}, showing one-to-one feasibly neat correspondence with the experimental data.  The calculated states are overpredicted up to 1.174 MeV experimental energy value and under predicted above this energy range for both the parities. 
In our calculation, the experimental level at 0.511 MeV is reproduced with large discrepancies 415 keV and 945 keV for $5/2_2^-$ and $3/2_4^-$ states, respectively, while $1/2_2^-$ is reproduced with 122 keV energy difference with respect to the experimental data.
We suggest that this experimental level can be associated with either the $1/2_2^-$ or $5/2_2^-$ calculated state.
For the experimental level at 0.867 MeV, the calculated state can be associated with  $7/2_2^-$ or $9/2_2^-$. 
Above $\sim$ 1 MeV  energy value, all the calculated states are compressed. 
{\color{black}The calculated $27/2_1^+$ state is lower in energy by 1 MeV with respect to the experimental data.}
The compression in energy in our calculated spectrum for high-spin states might be due to the need for core-excitation and significant configuration mixing with the higher orbitals beyond $Z=82$ and $N=126$ shell closure. As the neutron number increases in odd Rn isotopes, the calculated ${{1/2}_1}^-$, ${{3/2}_1}^-$ and ${{5/2}_1}^-$ states show strong single-particle nature with the dominant configuration $\nu(p_{1/2})^{-1}$ ($38\%$), $\nu(p_{3/2})^{-1}$ ($36\%$) and $\nu(f_{5/2})^{-1}$ ($39\%$), respectively. 

The shell-model energy spectrum of $^{211}$Rn in comparison with the experimental data is shown in Fig. \ref{fig8},  where all the observed levels up to 3.9 MeV excitation energy are reported. Experimentally, the relative energies between the ground state and many excited states are unknown in $^{211}$Rn. The $(17/2_1^-)$ state of these excited states is the lowest and observed at 1.578+x MeV. In this figure, these states are shown without any assumption of x ( i.e., x=0). Most of the levels in the case of $^{211}$Rn are tentative. For $^{211}$Rn, the shell-model spectrum reproduces very well the experimental data with an establishment of one-to-one correspondence in both positive and negative parity states, with good estimation for the tentative states. The calculated $5/2_1^-$ state is 147 keV higher than the experimental data, similar to other odd Rn isotopes for the first excited state. The tentative experimental state at 0.833 MeV is reproduced with only 31 keV energy difference in our calculation, suggesting this state as $3/2_1^-$. 
In the case of positive parity, the calculated $9/2_1^+$ is over-predicted, while $31/2_1^+$ and $35/2_1^+$ states are  lower in energies. This may be due to our model space is not enough for these high-spin states.
For the negative parity, the calculated ${{1/2}_1}^-$, ${{3/2}_1}^-$ and ${{5/2}_1}^-$ states show strong single-particle nature with the dominant configuration $\nu(p_{1/2})^{-1}$($48\%$), $\nu(p_{3/2})^{-1}$($42\%$) and $\nu(f_{5/2})^{-1}$($49\%$), respectively. 
Due to the restrictions of orbitals in the KSHELL code, as mentioned before, the KHH7B interaction could not reproduce high-lying states.

The shell-model energy spectrum of $^{213}$Rn in comparison with the experimental data is shown in Fig. \ref{fig9} with two interactions KHPE and KHH7B,  where all the observed levels up to 3.5 MeV excitation energy are reported. Experimentally, the relative energies between the ground state and many excited states are unknown in $^{213}$Rn. The $(25/2_1^+)$ state of these excited states is the lowest and observed at 1.664+x MeV. In this figure, these states are shown without any assumption of x ( i.e., x=0). 
In $^{213}$Rn, all levels are tentative, and many of the energy levels are unidentified in terms of the spin-parity. These unidentified levels are not included in the figure. Our calculation is supporting almost all of the tentative experimental states from both interactions. 
The shell-model results from the KHPE interaction are better than the KHH7B interaction for positive parity states, and the KHH7B interaction is slightly better for the negative parity states, as we can see in Fig. \ref{fig9}. 
This is due to the different model spaces in these interactions, which are responsible for generating the spectrum for positive or negative parity states. 
The $^{213}$Rn isotope is just above the $Z=82$ and $N=126$ shell closure. Therefore the trend of a higher first excited state, as in other odd Rn isotopes, is not followed here as it is compressed because of collectivity. For the energy value up to 2.121 MeV, KHPE interaction gives overall good results, and KHH7B interaction gives compressed results. 
Above this energy range, KHPE interaction slightly overpredicts the states, but overall results agree with the with the experimental data, and KHH7B interaction gives highly compressed results.
These compressed results from the KHH7B interaction indicate the importance of inclusion of the higher orbitals and core-excitation beyond $Z=82$ and $N=126$ shell closure. 
The KHPE interaction results for high-lying states are overpredicted because we have not taken into account core-excitation.  
The states above 3.0 MeV are compressed from the KHPE interaction. The KHH7B interaction has reproduced quite well the spectrum for the negative parity states. We can see from Fig. \ref{fig9}, our calculation can not predict precisely the unconfirmed ${27/2}$ experimental state, as this state is equally likely to be assigned in both the parities. 

The shell-model energy spectrum of $^{215}$Rn in comparison with the experimental data is shown in Fig. \ref{fig10} with the two interactions KHPE and KHH7B,  where all the observed levels up to 2.3 MeV excitation energy are reported. In $^{215}$Rn, all levels are tentative except for the ground state. The shell-model result is supporting a few of the tentative experimental states from both the interactions.
The first excited state is not confirmed experimentally. In our calculation, the ${{11/2}_1}^+$ state is coming as the first-excited state from both the interaction, which is an experimentally tentative second excited state. 
The calculated ${{7/2}_1}^+$ state has 119 keV and 246 keV difference and the ${{9/2}_1}^+$ state has 689 keV and 563 keV difference with the experimental value from KHPE and KHH7B interaction, respectively. 
Therefore, for the experimental state at 0.214 MeV, our calculation suggests the state as ${{7/2}_1}^+$.
For the experimental negative parity state at 0.291 MeV, our calculation overpredicts all three tentative spins with large energy value differences from the experimental data from both the interactions.


In all the Rn isotopes, we have observed that most of the low-lying states are collective because they show large configuration mixing. Some of these states are highly admixture of different configurations that have almost the same probability. On the other hand, the high-lying states are less collective, and many states in $^{212}$Rn are almost pure.

\begin{table*}
	\begin{center}
		\caption{The calculated $B(E2)$ values in units of W.u. for Rn
			isotopes using KHH7B interaction (SM) in comparison with the experimental data (Expt.) 
			\cite{NNDC,207Rn,208Rn,209Rn,210Rn,211Rn,212Rn,213Rn,214Rn,215Rn,216Rn} corresponding to $e_p$ = 1.5$e$ and $e_n$ = 0.5$e$.  }
		\label{t_be2}
		
		
		\begin{tabular}{rrc|cccc}
			\hline       
			\hline
			& ${B(E2; J_i \rightarrow   J_f}$)  & \hspace{1.0cm}~~~~~ &  \hspace{1.0cm}~~~~~ ${B(E2; J_i \rightarrow   J_f}$) &      &\\
			\hline
			\hline
			& &    &   &     &    \\
			$^{207}$Rn & Expt. & SM & $^{208}$Rn  & Expt. & SM     \\
			& &    &   &     &    \\
			\hline 
			$3/2_1^- \rightarrow   5/2_1^-$  & NA      & 1.428 & $2_1^+ \rightarrow  0_1^+$   &  NA       & 5.838       \\ 
			$9/2_1^- \rightarrow   5/2_1^-$ & NA & 5.551  & $4_1^+ \rightarrow  2_1^+$  & 4.7(4)  & 0.1789  \\ 
			$13/2_1^- \rightarrow   9/2_1^-$   & NA      & 1.347   &   $6_1^+ \rightarrow  4_1^+$ & NA       & 0.124     \\ 
			$17/2_1^+ \rightarrow   13/2_1^+$   &  NA     & 7.655  &   $8_1^+ \rightarrow  6_1^+$ & 0.187(7)       & 0.449  \\ 
			$17/2_1^- \rightarrow   13/2_1^-$   &  NA     & 0.511  &   $10_1^+ \rightarrow  8_1^+$ & NA       & 4.557  \\
			$21/2_1^+ \rightarrow   17/2_1^+$   &  NA     & 0.144  &   $14_1^+ \rightarrow  12_1^+$ & 0.35(17)       & 5.177  \\  
			
			\hline
			& &    &   &     &    \\
			$^{209}$Rn  & Expt. & SM  & $^{210}$Rn & Expt. & SM  \\
			& &    &   &     &    \\
			\hline 
			$1/2_1^- \rightarrow   5/2_1^-$ & NA &  0.808  & $2_1^+ \rightarrow  0_1^+$  & NA     &   5.286 \\ 
			$3/2_1^- \rightarrow   5/2_1^-$ & NA & 0.325  & $4_1^+ \rightarrow  2_1^+$  & 1.8(2) &  0.698 \\ 
			$9/2_1^- \rightarrow   5/2_1^-$ & NA     &  6.994 & $6_1^+ \rightarrow  4_1^+$   & 1.58(15)    &  0.963   \\  
			$13/2_1^- \rightarrow  9/2_1^-$ & NA     &   0.159   & $6_1^+ \rightarrow  4_2^+$   & 1.58(19)    &   1.752    \\ 
			$23/2_1^- \rightarrow  19/2_1^-$ & NA     &  1.954    & $8_1^+ \rightarrow  6_1^+$   & NA    &  0.351     \\ 
			$25/2_1^- \rightarrow  21/2_1^-$ & NA     &   5.028   & $10_1^+ \rightarrow  8_1^+$   & NA    &   3.937    \\ 
			$27/2_1^- \rightarrow  23/2_1^-$ & NA     &  1.774    & $12_1^+ \rightarrow  10_1^+$   & NA   &   2.734    \\ 
			$29/2_1^- \rightarrow  27/2_1^-$ & 0.66(15)     &  0.70    & $14_1^+ \rightarrow  12_1^+$   &  0.0248(23)   &  3.981$\times10^{-3}$     \\

			\hline
			& &    &   &     &    \\
			$^{211}$Rn  & Expt. & SM  & $^{212}$Rn & Expt. & SM  \\
			& &    &   &     &    \\ 
			\hline 
			$5/2_1^- \rightarrow   1/2_1^-$ & $>$0.040 & 1.911   & $2_1^+ \rightarrow  0_1^+$  & NA     &  5.195  \\ 
			$9/2_1^- \rightarrow   5/2_1^-$ & NA & 0.222  & $4_1^+ \rightarrow  2_1^+$  & 1.05$^{+44}_{-40}$ & 1.234  \\ 
			$13/2_1^- \rightarrow   9/2_1^-$ & NA     & 0.744  & $6_1^+ \rightarrow  4_1^+$   & 0.40$^{+6}_{-4}$    &  0.717   \\  
			$21/2_1^- \rightarrow   17/2_1^-$ & $>$0.030    &   1.744   & $8_1^+ \rightarrow  6_1^+$   & 0.117(7)    &   0.229    \\ 
			$25/2_1^- \rightarrow   21/2_1^-$ & $>$0.036     &   0.926   & $10_1^+ \rightarrow  8_1^+$   & NA    &  3.294     \\ 
			$29/2_1^- \rightarrow   25/2_1^-$ & 0.073(17)     &  0.012    & $12_1^+ \rightarrow  10_1^+$   & 4.52$^{+32}_{-29}$    &  3.115     \\ 
			$29/2_1^- \rightarrow   25/2_2^-$ &  1.9(6)    &   0.090   & $14_1^+ \rightarrow  12_1^+$   &  0.0319$^{+45}_{-36}$  &  5.236$\times10^{-3}$     \\ 
			$31/2_1^+ \rightarrow   27/2_1^+$ & $>$0.0077     &  1.688    & $14_1^+ \rightarrow  12_2^+$   &  2.9(6)    &   2.624    \\ 
			$35/2_1^+ \rightarrow   31/2_1^+$ & 2.3(5)     &    2.484   & $17_1^- \rightarrow  15_1^-$   & 2.94$^{+17}_{-15}$    &   2.942    \\

			\hline
			& &    &   &     &    \\
			$^{213}$Rn  & Expt. & SM  & $^{214}$Rn & Expt. & SM  \\
			& &    &   &     &    \\
			\hline 
			$13/2_1^+ \rightarrow   9/2_1^+$ & NA     & 4.645  & $2_1^+ \rightarrow  0_1^+$  & $>$0.032     &  5.459  \\ 
			$17/2_1^+ \rightarrow   13/2_1^+$ & NA     & 2.255  & $4_1^+ \rightarrow  2_1^+$  & $>$0.28 & 5.069  \\ 
			$21/2_1^+ \rightarrow   17/2_1^+$ & 1.68(16)     & 1.421  & $6_1^+ \rightarrow  4_1^+$   & 3.8$^{+17}_{-9}$    &  2.026   \\  
			$37/2_1^+ \rightarrow   33/2_1^+$ &  0.12(7)   &  2.329  & $8_1^+ \rightarrow  6_1^+$   & 3.8$^{+3}_{-1}$    &    0.251   \\ 
			$37/2_1^+ \rightarrow   33/2_2^+$ &  4(3)    & 0.208  & $10_1^+ \rightarrow  8_1^+$   & 2.9(7)    &   2.905$\times10^{-3}$   \\
			$35/2_1^- \rightarrow   31/2_1^-$ & NA     &  3.338 & $12_1^+ \rightarrow  10_1^+$   & $>$0.0064    &    1.39$\times10^{-4}$    \\

			\hline
			& &    &   &     &    \\
			$^{215}$Rn  & Expt. & SM  & $^{216}$Rn & Expt. & SM  \\
			& &    &   &     &    \\
			\hline 
			$11/2^+ \rightarrow   9/2^+$ & NA & 0.022  & $2^+ \rightarrow  0^+$  & NA     &  9.740  \\ 
			$13/2^+ \rightarrow   9/2^+$ & NA     & 3.733  & $4^+ \rightarrow  2^+$  & NA & 15.237  \\ 
			$15/2^+ \rightarrow   11/2^+$ & NA     & 3.531 & $6^+ \rightarrow  4^+$   & NA    &  7.499   \\  
			$17/2^+ \rightarrow   13/2^+$ & NA     & 4.226  & $8^+ \rightarrow  6^+$   & NA    &   0.331    \\ 
			$19/2^+ \rightarrow   15/2^+$ & NA     & 0.344   &    &     &       \\ 
			$21/2^+ \rightarrow   17/2^+$ & NA     &  2.087 &  &     &       \\ 
			\hline
		\end{tabular}
	\end{center}
\end{table*}

\begin{table*}
	\begin{center}
		\caption{The calculated (with KHH7B) magnetic dipole moments $\mu$ in units of
			$\mu_N$ and electric quadrupole moments $Q$ in units of $e$b for Rn isotopes (SM) in comparison with the experimental data (Expt.) \cite{NNDC,207Rn,208Rn,209Rn,210Rn,211Rn,212Rn,213Rn,214Rn,215Rn,216Rn}. The effective charges are taken as $e_p$ = 1.5$e$ and $e_n$ = 0.5$e$ for quadrupole moment.  The gyromagnetic ratios for magnetic moments are taken as $g_l^\nu$ = 0.00, $g_l^\pi$ = 1.00 for orbital angular momenta, and $g_s^\nu$ = -3.826, $g_s^\pi$ = 5.586 for spin angular momenta.}
		\label{t_Q}

		\begin{tabular}{r|rccc|c|cccccccccccc}
			\hline \hline
			\multicolumn{2}{c}\text{$\mu(\mu_{N})$ }  &  \multicolumn{2}{c}\text{{Q(eb)}} & &   \multicolumn{2}{c}\text{$\mu(\mu_{N})$ }   & \multicolumn{2}{c}\text{{Q(eb)}} &&\\ \hline
			\hline	
			&   &  &   & & &  &  &  &  \\
			$^{207}$Rn & Expt.  & SM  & Expt.  & SM & $^{208}$Rn & Expt.  & SM  & Expt.  & SM  \\
			&   &  &   & & &  &  &  &  \\
			\hline
			
			$1/2_1^-$ & NA & +0.4477 & - &  -    & $2_1^+$ & NA & +0.6774 & NA  & -0.0531 \\
			$3/2_1^-$ & NA  & -0.8955 & NA & +0.1165     & $4_1^+$ & NA & +1.4301 & NA & +0.4231      \\
			$5/2_1^-$ & +0.816(9)  & +0.9602 & +0.220(22) & +0.1432     & $6_1^+$ & NA & +3.5437 & NA & -0.0819   \\
			$9/2_1^-$ &   N/A    & +1.6032 & NA &  +0.1561      & $8_1^+$ & +6.98(8) & +4.7403 & +0.39(5) & -0.3628  \\
			$13/2_1^+$ &   -0.903(3)    & -1.1908 & NA &  +0.7427      & $10_1^+$ & NA & +5.6742  & NA & -0.3598 \\
			$17/2_1^+$ &   N/A      & -0.4300  & NA &  +0.7005     & $10_1^-$ & +10.77(10) & +10.4988 & NA & -1.4264\\
			\hline

			&   &  &   & & &  &  &  &  \\
			$^{209}$Rn & Expt.  & SM  & Expt.  & SM & $^{210}$Rn & Expt.  & SM  & Expt.  & SM  \\
			&   &  &   & & &  &  &  &  \\
			\hline
			
			$1/2_1^-$ & NA & +0.6453 & -  &  -    & $2_1^+$ & NA  & +0.3804 & NA & +0.3602 \\
			$3/2_1^-$ & NA  & -1.2722  & NA &  +0.1773    & $4_1^+$ & NA  & +2.3803  & NA &  +0.1283     \\
			$5/2_1^-$ & +0.8388(4)  & +1.3379 & +0.31(3)  & +0.3104     & $6_1^+$ & NA  & +3.5861 & NA &  -0.0529  \\
			$9/2_1^-$ &  NA  & +2.0306 & NA &  +0.5191     & $8_1^+$ & +7.184(56) & +4.8116 & +0.31(4) & -0.3766  \\
			$13/2_1^+$ &  NA  & -1.7442 & NA &  +0.6068      & $14_1^+$ & +14.92(10) & +12.0143 & NA & -1.1421 \\
			&      &   &  &       & $20_1^+$ & +22.3 & +20.8586 & NA & -1.7096 \\
			&      &   &  &       & $11_1^-$ & +12.16(11) & +11.3682  & NA  & -1.0208 \\
			&      &   &  &       & $17_1^-$ & +17.88(9) & +15.0538 & +0.86(10)  & -1.3051 \\
			
			\hline

			&   &  &   & & &  &  &  &  \\
			$^{211}$Rn & Expt.  & SM  & Expt.  & SM & $^{212}$Rn & Expt.  & SM  & Expt.  & SM  \\
			&   &  &   & & &  &  &  &  \\	 
			\hline
			
			$1/2_1^-$ & +0.601(7) & +0.6420  & - &   -  & $2_1^+$& NA  & +1.2235 & NA & +0.1411\\
			$3/2_1^-$ & NA  & -1.6115 & NA &  +0.1729    & $4_1^+$ & +4.0(2) & +2.3630  & NA &  +0.1067    \\
			$5/2_1^-$ & NA  & +1.4613 & NA & +0.2598     & $6_1^+$ & +5.45(5) & +3.5227 & NA & -0.0553  \\
			$9/2_1^-$ &   N/A    & +3.1991  & NA & +0.1127       & $8_1^+$ & +7.15(2) & +4.6982 & NA & -0.3126  \\
			$17/2_1^-$ & +7.75(8) & +5.4127 & +0.18(2) & -0.2953       & $14_1^+$ & +15.0(4) & +11.9660 & NA  & -0.8948 \\
			$43/2_1^-$ &   +15.9(4)  & +21.3675 & NA &  -1.4060      & $17_1^-$ & +17.9(2) & +14.9328 & NA & -1.0475 \\
			$35/2_1^+$ &  +17.80(21) & +15.0728  & NA & -1.1457     & $19_1^-$ & NA & +22.3372 & NA & -0.9955 \\
			\hline

			&   &  &   & & &  &  &  &  \\
			$^{213}$Rn & Expt.  & SM  & Expt.  & SM & $^{214}$Rn & Expt.  & SM  & Expt.  & SM  \\
			&   &  &   & & &  &  &  &  \\
			\hline
			
			$9/2_1^+$ & NA  & -1.7661  & NA &  -0.4454   & $2_1^+$ & NA & +0.0252 & NA &+0.3038\\
			$11/2_1^+$ & NA  & +1.6386  & NA &  -0.5370    & $4_1^+$ & NA & -0.0127 & NA & +0.1857    \\
			$13/2_1^+$ & NA  & -0.4519  & NA & -0.2827     & $6_1^+$ & NA & +2.4739 & NA & -0.1133 \\
			$17/2_1^+$ & NA  & +0.7723  & NA &  -0.3534     & $8_1^+$ & NA & -1.4036 & NA & -0.3693\\
			$21/2_1^+$ & +4.73(11)  & +2.3722  & NA & -0.7344      & $10_1^+$ & NA & -0.0557 & NA & -0.9853 \\
			$25/2_1^+$ & +7.63(25)  & +4.9119  & NA  & -0.7076     & $12_1^+$& NA &  & NA &-0.2969\\
			$15/2_1^-$ & NA  & -1.6079  &NA  & -0.6745     & $11_1^-$ & NA & -2.9506  & NA & -0.8147\\
			$31/2_1^-$ & +9.90(8)  & +3.9558  &NA  &  -0.6248     & $12_1^-$  & NA & +0.0166 & NA & -1.0387 \\
			\hline

			&   &  &   & & &  &  &  &  \\
			$^{215}$Rn & Expt.  & SM  & Expt.  & SM & $^{216}$Rn & Expt.  & SM  & Expt.  & SM  \\
			&   &  &   & & &  &  &  &  \\
			\hline
			
			$7/2_1^+$ &   N/A  & -1.3756 & NA & -0.5900    & $2_1^+$ & NA & &NA &+0.5094\\
			$9/2_1^+$ &   N/A & -1.7551 & NA &  -0.3608    & $4_1^+$ & NA & &NA &  +0.5465   \\
			$11/2_1^+$ &   N/A & +1.5711  & NA &  -0.9984    & $6_1^+$& NA & +1.4068 &NA & -0.0274 \\
			$13/2_1^+$ &   N/A & -1.0438 & NA &  -0.1496    & $8_1^+$ & NA & +4.7956 &NA & -0.4624 \\
			$15/2_1^+$ &   N/A & +1.7524 & NA &  -0.4138    & $10_1^+$& NA &-0.0161 &NA & -0.9063 \\
			$17/2_1^+$ &   N/A & -1.7198 & NA &  -0.1408    & $12_1^+$& NA & +1.8390 &NA & -0.5366 \\
			$7/2_1^-$ &   N/A & +0.9056 & NA &  -0.1517    & $13_1^-$& NA & -0.0292 &NA & -1.6812\\
			$9/2_1^-$ &   N/A & +0.8498 & NA &  +0.0961    & $15_1^-$ & NA & +1.2423 &NA & -1.1205 \\
			$11/2_1^-$ &   N/A & -0.9523 & NA &  -0.0738    &  &  & & & \\
			\hline \hline
			
		\end{tabular}
	\end{center}
\end{table*}

In $^{207}$Rn and $^{209}$Rn, the $13/2_1^+$ isomeric state is formed by purely $\nu(i_{13/2})^{-1}$ configuration, thus $\nu$ = 1. In $^{209}$Rn, the $29/2_1^-$[ $\pi(h_{9/2})^4\otimes\nu(f_{5/2})^{-1}$] isomeric state is formed by one pair breaking of the $\pi(h_{9/2})$ orbital and one unpaired neutron in $f_{5/2}$, thus $\nu$ = 3. This isomeric state is coming from $B(M1)$+$B(E2)$ transition, and our result for the half-life is 55.55 ns which is very close to the observed half-life. The $41/2_1^-$ isomeric state in our calculation is coming from the configuration $\pi(h_{9/2})^2(i_{13/2})^2\otimes\nu(f_{5/2})^{-1}$ with 49.48$\%$ probability, and also from the configuration $\pi(h_{9/2})^2(i_{13/2})^2\otimes\nu(p_{1/2})^{-1}$ with 17.27$\%$ probability. This isomeric state is formed by one pair breaking in $\pi(h_{9/2})$ and $\pi(i_{13/2})$ orbital each, and one unpaired neutron in $f_{5/2}$ or $p_{1/2}$, thus the seniority is $\nu$ = 5. The ${35/2_1}^+$ isomeric state is arising from the configuration $\pi(h_{9/2})^3(i_{13/2})\otimes\nu(f_{5/2})^{-1}$ with 58.53$\%$, and also from the configuration $\pi(h_{9/2})^3(i_{13/2})\otimes\nu(p_{1/2})^{-1}$ with 11.45$\%$ probability. In $^{211}$Rn, isomeric state  ${35/2_1}^+$ is arising from the same configuration $\pi(h_{9/2})^3(i_{13/2})\otimes\nu(p_{1/2})^{-1}$ as $^{209}$Rn with an increased probability of 85.45$\%$. This ${35/2_1}^+$ isomeric state is formed by one pair breaking in $\pi(h_{9/2})$ orbital and one unpaired proton in $h_{9/2}$ and $i_{13/2}$ orbital each and one unpaired neutron in $f_{5/2}$ or $p_{1/2}$ orbital, thus the seniority is $\nu$ = 5. In $^{211}$Rn, the calculated half-life for ${35/2_1}^+$ isomeric state is 44.74 ns which is very close to the experimental half-life 40.2(14) ns because  the  $B(E2)$ value and corresponding energy difference for this state are well reproduced in our calculation with respect to the experimental data. The ${17/2_1}^-$[$\pi(h_{9/2})^4\otimes\nu(p_{1/2})^{-1}$] isomeric state is formed by one pair breaking in $\pi(h_{9/2})$ orbital and one unpaired neutron in $p_{1/2}$ with seniority three ($\nu$ = 3). In our calculation, the $43/2_1^-$ isomeric state is arising from the configuration $\pi(h_{9/2})^2(i_{13/2})^2\otimes\nu(f_{5/2})^{-1}$, with 91.30$\%$ probability. This state is formed with one pair breaking in both the orbitals $\pi(h_{9/2})$ and $\pi(i_{13/2})$ and one unpaired neutron in $f_{5/2}$ orbital, thus the seniority is $\nu$ = 5. The isomeric state $49/2_1^+$ is arising from the configuration $\pi(h_{9/2})^2(i_{13/2})^2\otimes\nu(g_{9/2})$, with 61.78$\%$ probability. This state is formed with one pair breaking in both the  $\pi(h_{9/2})$ and $\pi(i_{13/2})$ orbitals, and one neutron in $g_{9/2}$ orbital, thus the seniority is $\nu$ = 5. The $63/2_1^-$ isomeric state is coming from the configuration $\pi(h_{9/2})^2(i_{13/2})^2\otimes\nu(g_{9/2})(i_{11/2})(f_{5/2})^{-1}$, with 88.04$\%$ probability. This isomeric state is formed by one pair breaking in $\pi(h_{9/2})$ and $\pi(i_{13/2})$ orbital each, and one neutron in both the orbitals $g_{9/2}$ and $i_{11/2}$, with one unpaired neutron in $f_{5/2}$ orbital, thus the seniority is $\nu$ = 7.

\begin{table*}
	\begin{center}
		\caption{The calculated half-life for Rn isotopes (SM) in comparison with the experimental data (Expt.) \cite{NNDC,207Rn,208Rn,209Rn,210Rn,211Rn,212Rn,213Rn,214Rn,215Rn,216Rn}.}
		\label{t_hl}

		\begin{tabular}{r|rccccccccccccccc}
			\hline 
			\hline	
			&   &  &   & & &  &  &  &  \\
			$J^{\pi}$ & $E_{\gamma}$  & $B(E\lambda)$ or   &  $B(E\lambda)$  & $B(M\lambda)$ & Expt. & SM \\
			& (MeV)  & $B(M\lambda) $  &  ($e^2$fm$^{2\lambda}$) & ($\mu_N^2$fm$^{2\lambda-2}$)&  T$_{1/2}$ & T$_{1/2}$  \\
			&   &  &   & & &  &  &  &  \\
			\hline
			\hline
			
			&   &  &   & & &  &  &  &  \\
			$^{208}$Rn  & &  &  &   & &   \\
			$8_1^+$ & 6$\times10^{-3}$ & $B(E2)$ & 32.90 &   & 487(12) ns & 1076 ns  \\
			$10_1^-$ & 0.296  & $B(M1)$ &  & 2$\times10^{-4}$ & 11.8(7) ns & 4.69 ns      \\
			$16_1^-$ & 0.143 & $B(M1)$ &  & 0.0819 & 18.3(4) ns & 0.03 ns    \\\hline
			
			&   &  &   & & &  &  &  &  \\
			$^{209}$Rn  & &  &  &   & &   \\
			$29/2_1^-$ &  2.29$\times10^{-3}$ & $B(M1)$+$B(E2)$ & 6$\times10^{7}$ &  1.09$\times10^{4}$  & 13.9(21) ns & 55.55 ns   \\
			\hline 
			
			&   &  &   & & &  &  &  &  \\
			$^{210}$Rn  & &  &  &   & &   \\
			$8_1^+$ & 0.019 & $B(E2)$ & 2.55$\times10^{4}$ &  & 644(40) ns & 344 ns  \\
			$14_1^+$ & 0.274 & $B(E2)$ & 1.82$\times10^{-1}$ &  & 76(7) ns & 1049 ns  \\
			$23_1^+$ & 0.843 & $B(E2)$ & 1.14$\times10^{-2}$ &  & 1.04(7) ns & 0.61 ns  \\
			$23_1^+$ & 0.445 & $B(E2)$ & 4.67$\times10^{-2}$ &  & 1.04(7) ns & 13.04 ns  \\\hline

			&   &  &   & & &  &  &  &  \\
			$^{211}$Rn  & &  &  &   & &   \\
			$17/2_1^-$ &  0.015 & $B(E2)$ & 20.72 &  & 596 (28) ns & 1660 ns   \\
			$35/2_1^+$ & 0.043 & $B(E2)$ & 185.39 &  & 40.2(14) ns & 44.74 ns   \\\hline

			&   &  &   & & &  &  &  &  \\
			$^{212}$Rn  & &  &  &   & &   \\
			$6_1^+$ & 0.093 & $B(E2)$ & 53.91 &  & 118(14) ns & 122 ns  \\
			$8_1^+$ & 8$\times10^{-3}$ & $B(E2)$ & 17.24 & & 0.91(3) $\mu$s & 2.04 $\mu$s     \\
			$14_1^+$ & 0.273 & $B(E2)$ & 0.40 &  & 7.4(9) ns & 800 ns  \\
			$17_1^-$ & 2$\times10^{-3}$ & $B(E2)$ & 187.04 & & 28.9(14) ns & 807 ns  \\\hline 
			
			&   &  &   & & &  &  &  &  \\
			$^{213}$Rn  & &  &  &   & &   \\
			$15/2_1^-$ & 0.478 & $B(M2)$ & & 0.28$\times10^{3}$ & 26(1) ns &  4.84 ns  \\
			$15/2_1^-$ & 1.158 & $B(E3)$ & 0.66$\times10^{4}$ & & 26(1) ns & 64.79 ns   \\
			$21/2_1^+$ & 0.574 & $B(E3)$ & 0.24$\times10^{-1}$ & & 29(2) ns & 2.29 s   \\
			$21/2_1^+$ & 0.08 & $B(E2)$ & 115.81 & & 29(2) ns & 61.81 ns   \\
			$31/2_1^-$ & 0.269 & $B(E3)$ & 0.33$\times10^{4}$  & & 1.36(7) $\mu$s & 1559 $\mu$s   \\
			$31/2_1^-$ & 0.168 & $B(E3)$ & 0.17$\times10^{5}$  & & 1.36(7) $\mu$s & 1326 $\mu$s  \\
			$37/2_1^+$ & 0.100 & $B(E2)$ & 281.99 & & 26(1) ns & 122.6 ns   \\
			$37/2_1^+$ &  0.078 & $B(E2)$ & 19.58 & & 26(1) ns & 369.6 ns   \\
			$49/2_1^+$ &  0.789  & $B(E3)$ & 0.82$\times10^{-1}$  & & 12(1) ns & 7.50$\times10^{-2}$ s   \\\hline

			&   &  &   & & &  &  &  &  \\
			$^{214}$Rn  & &  &  &   & &   \\
			$18_1^+$ & 0.114 & $B(E2)$ & 5.7145 &   & 44(3) ns & 894 ns  \\\hline

		\end{tabular}
	\end{center}
\end{table*}

In $^{213}$Rn, the isomeric state $15/2_1^-$ is formed by purely $\nu(j_{15/2})$ configuration, thus seniority $\nu$ = 1. The $B(M2)$ and $B(E3)$ transitions are responsible for the isomeric nature of the $15/2_1^-$ state. From both transition values, our calculation is giving satisfactory results for the half-life. The isomeric states $21/2_1^+$ and $25/2_1^+$ are arising from the configuration $\pi(h_{9/2})^4\otimes\nu(g_{9/2})$, having seniority $\nu$ = 3 with one pair breaking in the $\pi(h_{9/2})$ orbital and one neutron in the $g_{9/2}$ orbital. The isomeric state $21/2_1^+$ is coming from both the transitions $B(E2)$ and $B(E3)$. The half-life obtained from the calculated $B(E2)$ value is satisfactory. Whereas the half-life results from $B(E3)$ value is coming in order of seconds, in contrast to the experimental half-life of 29(2) ns, this is because our calculation is not predicting $B(E3)$ value correctly.
The $31/2_1^-$[$\pi(h_{9/2})^3(i_{13/2})\otimes\nu(g_{9/2})$] isomeric state is formed by one proton in $h_{9/2}$, $i_{13/2}$ orbitals each and one neutron in $g_{9/2}$ orbital, thus $\nu$ = 3. The calculated half-lives for this isomeric state are reproduced in the same order $\mu$s as the experimental data 1.36(7) $\mu$s but with very high magnitude, because of the inaccurate prediction of $B(E3)$ value, and also the energy differences are compressed by half in comparison with the experimental data. The $37/2_1^+$[$\pi(h_{9/2})^3(f_{7/2})\otimes\nu(g_{9/2})$] isomeric state is formed by one pair breaking in  $\pi(h_{9/2})$ orbital, and one proton in $h_{9/2}$, $f_{7/2}$ orbitals each and one neutron in $g_{9/2}$ orbital, thus $\nu$ = 5. The half-life of ${37/2_1}^+$ state is in satisfactory agreement with the experimental data. Unlike $^{211}$Rn, the ${43/2_1}^-$ isomeric state in $^{213}$Rn is arising from the configuration $\pi(h_{9/2})^3(i_{13/2})\otimes\nu(g_{9/2})$. This isomeric state is formed by one pair breaking in $\pi(h_{9/2})$ orbital, and one proton in $h_{9/2}$, $i_{13/2}$ orbitals each and one neutron in $g_{9/2}$ orbital, thus $\nu$ = 5. The $49/2_1^+$ isomeric state is formed with the same configuration and seniority ($\nu$ = 5) as in $^{211}$Rn with increased probability of wave-function. The half-life for $49/2_1^+$ isomeric state was also produced in seconds because of the inaccurate $B(E3)$ value, while the experimental half-life is 12(1) ns. The isomeric state ${55/2_1}^+$[$\pi(h_{9/2})(i_{13/2})^3\otimes\nu(j_{15/2})$] is formed by one pair breaking in $\pi(i_{13/2})$ orbital, and one proton in $h_{9/2}$, $i_{13/2}$ orbitals each and one neutron in $j_{15/2}$ orbital, thus $\nu$ = 5.  In this way, we can see that for odd Rn isotopes $\pi(h_{9/2})$ and $\pi(i_{13/2})$ orbitals are responsible for forming most of the isomeric states.  
 On the whole, the calculated half-life values are in good agreement with the experimental data. 
 Previously, we have reported a shell-model study of isomeric states for $fp$ shell nuclei and Sn isotopes in Refs. \cite{fp, Sn}. 
In the $fp$ region $f_{7/2}$ and $g_{9/2}$ orbitals, in the Sn region $g_{7/2}$, $d_{5/2}$ and $h_{11/2}$ orbitals, while for the Pb region $h_{9/2}$, $f_{7/2}$ and $i_{13/2}$ orbitals are crucial. 
Our result corresponding to isomeric states for the Rn chain shows the importance of $h_{9/2}$, $f_{7/2}$ and $i_{13/2}$ orbitals.
 {\color{black}  Several recent articles are available in the literature to explain seniority isomer for different nuclei within the framework of the nuclear shell-mode 
  \cite{astier,PRC85astier3,PRC87astier2}.}

\subsection{Electromagnetic properties}

In this section, we have discussed the results of  the $B(E2)$ values, magnetic moments, and quadrupole moments for Rn isotopes.
For $^{207}$Rn, the experimental data for $B(E2)$ values are not available. 
 In $^{208}$Rn,  our calculated value for $B(E2; 4_1^+ \rightarrow 2_1^+)$ 
is smaller than the experimental data. This may be because the $2_1^+$ and $4_1^+$ states show a large configuration mixing
in our calculation. For $B(E2; {8_1}^+ \rightarrow {6_1}^+)$ transition, a small value is reproduced as the experimental data.
In $^{208}$Rn, because of this small $B(E2)$ value, the $8_1^+$ state is an isomer with 487 (12) ns half-life \cite{NNDC}. A large $B(E2; 14_1^+ \rightarrow 12_1^+)$  value 5.177 W.u. is obtained  in our calculation corresponding to a small experimental value 0.35(17) W.u. In theory, the ${14_1}^+$ and ${12_1}^+$ states consist of the same configuration [$\pi(h_{9/2})^4\nu(f_{5/2})^{-2}$]. This is why the theoretical $B(E2; 14_1^+ \rightarrow 12_1^+)$ value is large. The calculated magnetic and quadrupole moments
are in good agreement with the experimental data, although, the sign is different for quadrupole moment in $^{208}$Rn.
In $^{210}$Rn, the calculated $B(E2; 4_1^+ \rightarrow 2_1^+)$ value is almost half of the experimental value. In our calculation, the $2_1^+$ and $4_1^+$  states come from the same configuration [$\pi(h_{9/2})^4$], but the $2_1^+$ state shows a large configuration mixing. This is the reason for the small $B(E2)$ value. Corresponding to the experimental $B(E2; 14_1^+ \rightarrow 12_1^+)$ value 0.0248(23) W.u., our calculation reproduces a very small value which is 3.981$\times10^{-3}$ W.u. The $14_1^+$ state is with configuration [$\pi(h_{9/2})^3(f_{7/2})$] and the $12_1^+$ state is with configuration [$\pi(h_{9/2})^4$]. This difference in the configurations is responsible for the small $B(E2)$ value. This hindered  decay shows the behaviour of the isomeric state $14_1^+$ in the $^{210}$Rn with 76(7) ns half-life \cite{NNDC}.

In $^{211}$Rn, overall calculated $B(E2)$ values are in  good agreement with the experimental data. 
The calculated magnetic and quadrupole moments are well reproduced with the experimental data, although the sign is different for the quadrupole moment of $17/2_1^-$ state. 
In $^{212}$Rn, most of the $B(E2)$ values are well reproduced in our calculation. The largest discrepancy between the experimental and calculated  value is seen in the  $B(E2; 14_1^+ \rightarrow 12_1^+)$ transition. In our calculation, the $12_1^+$ and $12_2^+$ states are coming from the configurations [$\pi(h_{9/2})^4$] and [$\pi(h_{9/2})^3(f_{7/2})$], respectively. Whereas $14_1^+$ state is coming from the [$\pi(h_{9/2})^3(f_{7/2})$] configuration. Due to this difference of the configurations, the small value is reproduced for $B(E2; 14_1^+ \rightarrow 12_1^+)$ transition, and because of the same configuration, the calculated $B(E2; 14_1^+ \rightarrow 12_2^+)$ value is large.  In $^{212}$Rn, because of this hindered decay,
our calculation supports the isomeric nature of $14_1^+$ state. The experimental half-life of this
state is  7.4(9) ns \cite{NNDC}, similar to the $14_1^+$  isomeric state in $^{210}$Rn.
In $^{213}$Rn, the experimental $B(E2; 37/2_1^+ \rightarrow 33/2_1^+)$ and  $B(E2; 37/2_1^+ \rightarrow 33/2_2^+)$ values are 0.12(7) W.u. and 4(3) W.u., respectively. The $33/2_1^+$ and $33/2_2^+$ states are calculated higher than the $37/2_1^+$ state in contrast to the experimental spectrum. 
The calculated $B(E2)$ value corresponding to $B(E2; 37/2_1^+ \rightarrow 33/2_1^+)$,
$B(E2; 37/2_1^+ \rightarrow 33/2_2^+)$ transitions are  2.329 and 0.208 W.u., respectively.  In our calculation, $37/2_1^+$ and $33/2_1^+$ states are arising from the configuration [$\pi(h_{9/2})^3(f_{7/2})\nu(g_{9/2})$], hence giving a large $B(E2)$ value. In contrast, the calculated $B(E2; 37/2_1^+ \rightarrow 33/2_2^+)$ value is small as the $33/2_2^+$ state is coming from a different configuration [$\pi(h_{9/2})^4\nu(g_{9/2})$].   Therefore, the $33/2_1^+$ and $33/2_2^+$ states might be reversely calculated referring to the $37/2_1^+$ state, compared to the experimental data.
In $^{214}$Rn, the calculation predicts large $B(E2)$ values for $B(E2; {2_1}^+ \rightarrow {0_1}^+)$  and $B(E2; {4_1}^+ \rightarrow {2_1}^+)$ transitions, whereas smaller $B(E2)$ values for $B(E2; 8_1^+ \rightarrow 6_1^+)$, $B(E2; 10_1^+ \rightarrow 8_1^+)$ and $B(E2; 12_1^+ \rightarrow 10_1^+)$ transitions. In $^{214}$Rn, these states are coming from huge configuration mixing. On the whole, theoretical calculations for the electromagnetic properties reproduce the experimental data well. 
The $B(E2)$ values are small, corresponding to the transitions from the isomeric states with the shell-model.
We have also reported electromagnetic properties for different states where experimental data
are not available. Since our calculation is providing overall good agreement with the experimental data,  our prediction might be helpful for a future experiment.

\section{CONCLUSIONS}\label{4} 

 In the present study, we have performed systematic shell-model calculations for the  $^{207-216}$Rn isotopes employing two different interactions KHH7B and KHPE, developed for the different model spaces.
For $^{207-212}$Rn isotopes, we have used KHH7B interaction, while for $^{213-216}$Rn, we have performed calculations with KHH7B and KHPE interactions.
The results of the KHH7B interaction are in reasonable agreement with the experimental data. The KHPE interaction is found to give a good description for the  $^{213-216}$Rn isotopes. The cross-shell interaction KHH7B is found to give good agreements with the experimental data for the isotopes below $N=126$ shell-gap.
For higher mass isotopes $^{213-216}$Rn, it becomes crucial to consider sufficient orbitals below the shell-closure for low-lying states apart from the core-excitation.
We have successfully reproduced all the new levels identified in the experiment for  $^{212}$Rn \cite{212Rn2}. Our calculation also supports many tentative levels in the energy spectrum for other Rn isotopes. We found that low-lying states are arising from large configuration mixing while high-lying states show less collective behavior. 
 The isotopes near $N=126$ shell closure show minor collective nature, and as we move far from the shell-closure, collectivity and configuration mixing increase very rapidly.
This shows the importance of the inclusion of a sufficient model space in the $^{208}$Pb region to reproduce the energy spectrum correctly.
We have calculated $B(E2)$ values, magnetic and quadrupole moments, and compared  with the available experimental data. We have also reported shell model results where the experimental data are not available. This will be very useful to compare the upcoming experimental data. 

We have also analyzed different isomeric states and calculated corresponding half-lives. The calculated $B(E2)$ value supports the behavior of these isomeric states so that the half-lives of these isomeric states are well reproduced.
The isomeric states are described in terms of the shell-model configuration and seniority quantum number ($v$).  The orbitals $h_{9/2}$ and $i_{13/2}$ 
are responsible for the isomeric states in the Rn isotopes.
The high-spin isomers in Rn isotopes are due to seniority   ($v$) = 1, 2, 3, 4, 5, 6 and 7.

\section*{ACKNOWLEDGEMENTS}
We acknowledge financial support from MHRD, the Government
of India,  and a research grant from SERB (India), CRG/2019/000556.
Shell-model calculations were performed at the Kalam and Prayag computational facilities
at IIT-Roorkee. We would like to thank Prof. Larry Zamick for useful discussions.

}


\begin{thebibliography}{44}
	\expandafter\ifx\csname natexlab\endcsname\relax\def\natexlab#1{#1}\fi
	\expandafter\ifx\csname bibnamefont\endcsname\relax
	\def\bibnamefont#1{#1}\fi
	\expandafter\ifx\csname bibfnamefont\endcsname\relax
	\def\bibfnamefont#1{#1}\fi
	\expandafter\ifx\csname citenamefont\endcsname\relax
	\def\citenamefont#1{#1}\fi
	\expandafter\ifx\csname url\endcsname\relax
	\def\url#1{\texttt{#1}}\fi
	\expandafter\ifx\csname urlprefix\endcsname\relax\def\urlprefix{URL }\fi
	\providecommand{\bibinfo}[2]{#2}
	\providecommand{\eprint}[2][]{\url{#2}}
	
\bibitem{Brown2000}
B. A. Brown,
Double-octupole states in $^{208}$Pb, 
\href{https://doi.org/10.1103/PhysRevLett.85.5300}
{\color{blue} Phys. Rev. Lett. {\bf 85}, 5300 (2000)}.


\bibitem{Butler}	
P.A. Butler  \textit{et al.},
Evolution of Octupole Deformation in Radium Nuclei from Coulomb Excitation of Radioactive $^{222}$Ra and $^{228}$Ra Beams,
\href{https://doi.org/10.1103/PhysRevLett.124.042503}
{\color{blue} Phys. Rev. Lett.  {\bf 124}, 042503 (2020)}.

\bibitem{T.Otsuka} T. Otsuka and Y. Tsunoda, 
The role of shell evolution in shape coexistence, 
\href{https://doi.org/10.1088/0954-3899/43/2/024009}
{\color{blue} J. Phys. G: Nucl. Part. Phys. {\bf 43},  024009 (2016).}

\bibitem{yosi} N. Yoshinaga, K. Yanase,  K. Higashiyama,  and E. Teruya, 
Octupole phonon model based on the shell model for octupole vibrational states, 
\href{https://doi.org/10.1103/PhysRevC.98.044321}
{ \color{blue} Phys. Rev. C {\bf 98}, 044321 (2018).}

\bibitem{Tang}
T.L. Tang \textit{et al.}, 
First Exploration of Neutron Shell Structure below Lead and beyond $N=126$,
\href{https://doi.org/10.1103/physrevlett.124.062502}
{ \color{blue}Phys. Rev. Lett.  {\bf 124}, 062502 (2020)}.

\bibitem{210At}S. Bayer \textit{et al.},
Core-excited states and core-polarization effects in $^{210}$At and $^{211}$At,
\href{https://doi.org/10.1016/S0375-9474(01)00922-8}
{\color{blue} Nucl. Phys. A {\bf 694}, 3 (2001)}.




\bibitem{210Ra} J. J. Ressler \textit{et al.}, 
Isomer decay tagging in the heavy nuclei: $^{210}\mathrm{Ra}$ and $^{209}\mathrm{Ra}$,
\href {https://link.aps.org/doi/10.1103/PhysRevC.69.034331}
{ \color{blue}Phys. Rev. C {\bf 69}, 034331 (2004)}.

\bibitem{Ra196} F. Hessberger, S. Hofmann, I. Kojouharov, and D. Ackermann,
Decay properties of isomeric states in radium isotopes close to $N = 126$,
\href{https://doi.org/10.1140/epja/i2003-10233-9}
{ \color{blue}Eur. Phys. J. A {\bf 22}, 253 (2004)}.

\bibitem{210Po} G. D. Dracoulis \textit{et al.},
Neutron core excitations in the $N=126$ nuclide $^{210}\mathrm{Po}$,
\href{https://link.aps.org/doi/10.1103/PhysRevC.77.034308}
{\color{blue}  Phys. Rev. C {\bf 77}, 034308 (2008)}.

\bibitem{208Fr} G. D. Dracoulis \textit{et al.},
Assignment of levels in $^{208}$Fr and $10^-$ isomers in the odd-odd isotones $^{206}$At and $^{208}$Fr,
\href{https://doi.org/10.1140/epja/i2009-10756-y}
{ \color{blue} Eur. Phys. J. A {\bf 40},  127 (2009)}.

\bibitem{206Bi} N. Cieplicka \textit{et al.}, 
Yrast structure of ${}^{206}$Bi: Isomeric states and one-proton-particle, three-neutron-hole excitations,
\href{https://link.aps.org/doi/10.1103/PhysRevC.86.054322}
{ \color{blue}  Phys. Rev. C {\bf 86}, 054322 (2012)}.



\bibitem{Discovery1} C. Fry and M. Thoennessen, 
Discovery of the thallium, lead, bismuth, and polonium isotopes,
\href{https://doi.org/10.1016/j.adt.2012.01.005}
{ \color{blue}  Atomic Data and Nuclear Data Tables {\bf 99}, 365 (2013)}.

\bibitem{Discovery2} C. Fry and M. Thoennessen,
Discovery of the astatine, radon, francium, and radium isotopes,
\href{https://doi.org/10.1016/j.adt.2012.05.003}
{ \color{blue} Atomic Data and Nuclear Data Tables {\bf 99}, 497 (2013)}.

\bibitem{prgati} Prgati \textit{et al.}, 
Parity doublet structures in doubly-odd $^{216}$Fr,
\href{https://doi.org/10.1103/PhysRevC.97.044309}
{ \color{blue} Phys. Rev. C {\bf 97}, 044309 (2018).}


\bibitem{berry} T.A. Berry \textit{et al.}, 
Octupole states in $^{207}\mathrm{Tl}$ studied through $\ensuremath{\beta}$ decay,
\href{https://link.aps.org/doi/10.1103/PhysRevC.101.054311}
{ \color{blue} Phys. Rev. C {\bf 101}, 054311 (2020)}.


\bibitem{abinitio}
T. Miyagi, S. R. Stroberg, P. Navr\'atil, K. Hebeler, and J. D. Holt,
Converged ab initio calculations of heavy nuclei,
{\color{blue} arXiv:2104.04688 }

\bibitem{Butlerjpg} P. A. Butler, J. Cederkall, and P. Reiter, 
Nuclear-structure studies of exotic nuclei
with MINIBALL, 
\href{https://doi.org/10.1088/1361-6471/aa5c4e}
{ \color{blue} Phys. G: Nucl. Part. Phys. {\bf 44},  044012 (2017).}



\bibitem{isacker} P. Van Isacker, 
A solvable model for octupole phonons,
\href{https://link.springer.com/10.1140/epjst/e2020-000026-x}
{\color{blue} Eur. Phys. J. Special Topics {\bf 229}, 2443 (2020)}.



\bibitem{jain} A. K. Jain \textit{et al.},
Atlas of Nuclear Isomers, 
\href{https://doi.org/10.1016/j.nds.2015.08.001}
{\color{blue}  Nucl. Data Sheets {\bf 128}, 1 (2015)}.

\bibitem{astier} A. Astier and M-G. Porquet,
First proton-pair breaking in semi-magic nuclei beyond ${}^{\mathbf{132}}$Sn and ${}^{\mathbf{208}}$Pb: Configuration of the long-lived isomer of ${}^{\mathbf{217}}$Pa,
\href{https://doi.org/10.1103/PhysRevC.87.014309}
{\color{blue} Phys. Rev. C {\bf 87}, 014309 (2013)}.

\bibitem{203Tl} 
V. Bothe, S.K. Tandel, S.G. Wahid, P.C. Srivastava, B. Bhoy, P. Chowdhury, R.V.F. Janssens, F.G. Kondev,
M.P. Carpenter, T. Lauritsen, D. Seweryniak, S. Zhu
{\color{blue} arXiv:2106.02314}.

\bibitem{phil}
P. Walker and Z. Podoly\'ak,
100 years of nuclear isomers—then and now,
\href{https://doi.org/10.1088/1402-4896/ab635d}
{\color{blue}  Phys. Scr. 95, 044004 (2020)}.


\bibitem{212Rn1} G. D. Dracoulis \textit{et al.},
Structure of the $N=126$ nuclide $^{212}\mathrm{Rn}$: Valence and core excited configurations,
\href{https://doi.org/10.1103/PhysRevC.80.054320}
{ \color{blue} Phys. Rev. C {\bf 80}, 054320 (2009)}.

\bibitem{212Rn2} C. B. Li \textit{et al.}, 
New level scheme and shell model description of $^{212}\mathrm{Rn}$,
\href{https://doi.org/10.1103/PhysRevC.101.044313}
{ \color{blue} Phys. Rev. C {\bf 101}, 044313 (2020)}.

\bibitem{Mcgrory} J. B. Mcgrory and T. T. S. Kuo,
Shell model calculations of two to four identical-“particle” systems near $^{208}$Pb,
\href{https://doi.org/10.1016/0375-9474(75)90637-5}
{ \color{blue}  Nucl. Phys. A {\bf 247}, 283 (1975)}.

\bibitem{Coraggio} L. Coraggio, A. Covello, A. Gargano, N. Itaco, and T. T. S. Kuo,
Bonn potential and shell-model calculations for $N=126$ isotones,
\href{https://link.aps.org/doi/10.1103/PhysRevC.60.064306}
{ \color{blue} Phys. Rev. C {\bf 60}, 064306 (1999)}.

\bibitem{Caurier} E. Caurier, M. Rejmund, and H.Grawe,
Large-scale shell model calculations for the $N=126$ isotones Po--Pu,
\href{https://link.aps.org/doi/10.1103/PhysRevC.67.054310}
{ \color{blue} Phys. Rev. C {\bf 67}, 054310 (2003)}.

\bibitem{koji} K. Higashiyama, N.  Yoshinaga,
Shell model description of low-lying states in Po and Rn isotopes,
\href{https://doi.org/10.1051/epjconf/20146602050}
{ \color{blue} EPJ Web of conferences {\bf 66}, 02050 (2014)}.

\bibitem{Teruya} E. Teruya, K. Higashiyama, N. Yoshinaga,
Large-scale shell-model calculations of nuclei around mass 210,
\href{https://link.aps.org/doi/10.1103/PhysRevC.93.064327}
{ \color{blue}Phys. Rev. C {\bf 93}, 064327 (2016)}.

\bibitem{Yanase} K. Yanase, E. Teruya, K. Higashiyama, N. Yoshinaga,
Shell-model study of Pb, Bi, Po, At, Rn, and Fr isotopes with masses from 210 to 217,
\href{https://link.aps.org/doi/10.1103/PhysRevC.98.014308}
{ \color{blue} Phys. Rev. C {\bf 98}, 014308 (2018)}.



\bibitem{Naidja} H. Naïdja, 
New shell-model investigation of the lead-208 mass region: Spectroscopic properties and collectivity,
\href{https://link.aps.org/doi/10.1103/PhysRevC.103.054303}
{ \color{blue} Rev. C {\bf 103}, 054303 (2021)}.



\bibitem{Wilson} E. Wilson \textit{et al.}, 
Core excitations across the neutron shell gap in $^{207}$Tl,
\href{http://dx.doi.org/10.1016/j.physletb.2015.04.055}
{\color{blue} Phys. Lett. B {\bf 747}, 88 (2015).}

\bibitem{Wahid} S.G. Wahid \textit{et al.}, 
Metastable states from multinucleon excitations in $^{202}\mathrm{Tl}$ and $^{203}\mathrm{Pb}$, 
\href{https://link.aps.org/doi/10.1103/PhysRevC.102.024329}
{ \color{blue} Phys.Rev. C {\bf 103}, 024329 (2020).}

\bibitem{Anil} A. Kumar and P.C. Srivastava, 
Shell-model description for the first-forbidden $\beta^-$ decay of $^{207}$Hg into the one-proton-hole nucleus $^{207}$Tl, 
\href{https://doi.org/10.1016/j.nuclphysa.2021.122255}
{ \color{blue} Nucl. Phys. A {\bf 1014}, 122255 (2021)}.


\bibitem{Nushellx1} B. A. Brown and W. D. M. Rae,
The Shell-Model Code NuShellX@MSU,
\href{https://doi.org/10.1016/j.nds.2014.07.022}
{\color{blue} Nucl. Data Sheets {\bf 120}, 115 (2014)}.

\bibitem{Nushellx2} B. A. Brown, 
The Nuclear Shell Model Towards the Drip Lines, 
\href{https://doi.org/10.1016/S0146-6410(01)00159-4} 
{ \color{blue} Prog. Part. Nucl. Phys. {\bf 47}, 517 (2001)}.


\bibitem{Kshell}N. Shimizu, T. Mizusaki, Y. Utsuno and Y. Tsunoda, 
Thick-restart block Lanczos method for large-scale shell-model calculations, 
\href{https://doi.org/10.1016/j.cpc.2019.06.011}
{ \color{blue}Comput. Phys. Comm. {\bf 244}, 372 (2019).}

\bibitem{Warburton1} E. K. Warburton and B. A. Brown, 
Appraisal of the Kuo-Herling shell-model interaction and application to $A = 210-212$ nuclei,
\href{https://link.aps.org/doi/10.1103/PhysRevC.43.602}
{ \color{blue} Phys. Rev. C {\bf 43}, 602 (1991)}.


\bibitem{pbpop} N. A. F. M. Poppelier and P. W. M. Glaudemans,
Particle-Hole Excitations in the $^{208}$Pb Mass Region, 
\href{https://link.springer.com/content/pdf/10.1007/BF01290233.pdf}
{  Z. Phys. A {\bf 329}, 275 (1988)}.

\bibitem{Kuo1} T. T. S. Kuo and G. Herling, US Naval Research Laboratory Report no. 2258 (1971) unpublished.

\bibitem{Kuo2} G. Herling and T. T. S. Kuo, 
Two-particle states in 210Pb, 210Bi and 210Po with realistic forces,
\href{https://doi.org/10.1016/0375-9474(72)90905-0}
{ \color{blue} Nucl. Phys. A {\bf 181}, 113 (1972)}.

\bibitem{Hamada} T. Hamada and I.D. Johnston, 
A potential model representation of two-nucleon data below 315 MeV,
\href{https://doi.org/10.1016/0029-5582(62)90228-6}
{\color{blue}  Nucl. Phys. {\bf 34}, 382 (1962)}.

\bibitem{Kuo3} T.T.S. Kuo and G.E. Brown,
Structure of finite nuclei and the free nucleon-nucleon interaction: An application to $^{18}$O and $^{18}$F,
\href{https://doi.org/10.1016/0029-5582(66)90131-3}
{\color{blue}  Nucl. Phys. {\bf 85}, 40 (1966)}.



\bibitem{Hosaka} A. Hosaka, K.-I. Kubo, H. Toki, 
$G$-matrix effective interaction with the paris potential,
\href{https://doi.org/10.1016/0375-9474(85)90292-1}
{ \color{blue} Nucl. Phys. A {\bf 444}, 76 (1985)}.


\bibitem{NNDC} Evaluated Nuclear Structure Data File (ENSDF)
\href{http://www.nndc.bnl.gov/ensdf/}
{\color{blue} http://www.nndc.bnl.gov/ensdf/}.

\bibitem{207Rn} F. G. Kondev and S. Lalkovski,
Nuclear Data Sheets for $A = 207$,
\href{https://doi.org/10.1016/j.nds.2011.02.002}
{ \color{blue} Nucl. Data Sheets {\bf 112}, 707 (2011)}.

\bibitem{208Rn} M. J. Martin,
Nuclear Data Sheets for $A = 208$,
\href{https://doi.org/10.1016/j.nds.2007.07.001}
{ \color{blue} Nucl. Data Sheets {\bf 108}, 1583 (2007)}.

\bibitem{209Rn} J. Chen and F. G. Kondev,
Nuclear Data Sheets for $A = 209$,
\href{https://doi.org/10.1016/j.nds.2015.05.003}
{ \color{blue}  Nucl. Data Sheets {\bf 126}, 373 (2015)}.

\bibitem{210Rn} M. S. Basunia, 
Nuclear Data Sheets for $A = 210$,
\href{https://doi.org/10.1016/j.nds.2014.09.004}
{ \color{blue} Nucl. Data Sheets {\bf 121}, 561 (2014)}.

\bibitem{211Rn} B. Singh {\it et al.}
Nuclear Data Sheets for $A = 211$,
\href{https://doi.org/10.1016/j.nds.2013.05.001}
{\color{blue} Nucl. Data Sheets {\bf 114}, 661 (2013)}.

\bibitem{212Rn} K. Auranen and E. A. McCutchan, 
Nuclear Data Sheets for $A=212$,
\href{https://doi.org/10.1016/j.nds.2020.09.002}
{ \color{blue} Nucl. Data Sheets {\bf 168}, 117 (2020)}.

\bibitem{213Rn} M. S. Basunia,
Nuclear Data Sheets for $A = 213$,
\href{https://doi.org/10.1016/j.nds.2007.02.002}
{ \color{blue} Nucl. Data Sheets {\bf 108}, 663 (2007)}.

\bibitem{214Rn} S.-C. Wu, 
Nuclear Data Sheets for A = 214,
\href{https://doi.org/10.1016/j.nds.2009.02.002}
{\color{blue}  Nucl. Data Sheets {\bf 110}, 681 (2009)}.

\bibitem{215Rn} B. Singh {\it et al.,}, 
Nuclear Data Sheets for $A = 215$,
\href{https://doi.org/10.1016/j.nds.2013.11.003}
{\color{blue}  Nucl. Data Sheets {\bf 114}, 2023 (2013)}.

\bibitem{216Rn} S.-C. Wu, 
Nuclear Data Sheets for A = 216,
\href{https://doi.org/10.1016/j.nds.2007.04.001}
{ \color{blue} Nucl. Data Sheets {\bf 108}, 1057 (2007)}.

\bibitem{fp} P.C. Srivastava and L. Zamick,
Minimal theory of isomerism -Q.Q and other  interactions, 
\href{https://doi.org/10.1088/1402-4896/abf306}
{\color{blue} Physica Scripta  {\bf 96}, 065307 (2021).}


\bibitem{Sn} P.C. Srivastava, B. Bhoy and M.J. Ermamatov,
Different seniority states of $^{119-126}$Sn isotopes: shell-model description, 
\href{https://doi.org/10.1093/ptep/ptz108}
{ \color{blue}Prog. Theor. Exp. Phys. {\bf 2019}, 103D01 (2019).}



\bibitem{PRC85astier3}
A. Astier et al.,High-spin states with seniority $v=4$, 5, and 6 in ${}^{119-126}$Sn,
\href{https://link.aps.org/doi/10.1103/PhysRevC.85.054316}
{ \color{blue} Phys. Rev. C {\bf 85}, 054316 (2012).}

 
\bibitem{PRC87astier2}
A. Astier et al., High-spin structures of $^{136}$Cs,
\href{https://link.aps.org/doi/10.1103/PhysRevC.87.054316}
{ \color{blue} Phys. Rev. C  {\bf 87}, 054316 (2013).}




 


\end{thebibliography}
\end{document}